\documentclass[12pt,preprint]{aastex}
\newcommand{\etal}{{\it et al. \,}}

\lefthead{Aparicio \etal}
\righthead{DDO 187}

\received{2007 May 22}
\begin{document}
\title{IAC-pop: finding the Star Formation History of resolved galaxies}

\author{Antonio Aparicio\altaffilmark{1}$^{\rm ,}$\altaffilmark{2}} 
\email{antapaj@iac.es}
\and
\author{Sebastian L. Hidalgo\altaffilmark{2}} 
\email{shidalgo@iac.es}

\altaffiltext{1}{Departamento de Astrof\'\i sica, Universidad de La Laguna. V\'\i a L\'actea s/n
E38200 - La Laguna, Tenerife, Canary Islands, Spain}
\altaffiltext{2}{Instituto de Astrof\'\i sica de Canarias. V\'\i a L\'actea s/n. 
E38200 - La Laguna, Tenerife, Canary Islands, Spain}

\begin{abstract}
  
  IAC-pop is a code designed to solve the star formation history (SFH)
  of a complex stellar population system, like a galaxy, from the analysis of
  the color-magnitude diagram (CMD). It uses a genetic
  algorithm to minimize a $\chi^2$ merit function comparing
  the star distributions in the observed CMD and the CMD of a synthetic
  stellar population. A parametrization of the CMDs is used, which is the main input of the code. In fact, the code
  can be applied to any problem in which a similar 
  parametrization of an experimental set of data and models can be made.
  
The method internal consistency and robustness against several error sources, including observational effects, data sampling and stellar evolution library differences, are tested. It is found that the best stability of the solution and the best way to estimate errors is obtained by several runs of IAC-pop with varying the input data parametrization. The routine {\it MinnIAC} is used to control this process. 
  
  IAC-pop is offered for free use and can be downloaded from the site {\tt
  http://iac-star.iac.es/iac-pop}. The routine MInnIAC is also offered under request, but support can not be provided for its use. The only requirement for the use of IAC-pop and MinnIAC is referencing this paper and crediting as indicated in the site. 

\end{abstract}


\section{Introduction \label{intr}}

Galaxies evolve on two main paths: dynamically, including interaction with
external systems, and through the process of formation, evolution and death of
stars within them. The latter has the following relevant effects on the
galaxy: (i) the evolution of gas content, (ii) the chemical enrichment and
(iii) the formation of the stellar populations with different properties as
the gas from which they form evolves. The star formation history (SFH) is
therefore fundamental to understand the galaxy evolution process.

The color-magnitude diagram (CMD) is, in practice, the best tool to study and derive the
star formation history of resolved galaxies. Deep enough CMDs display stars born
all over the life-time of the galaxy and are indeed fossil records of the
SFH. An approximate, qualitative sketch of the stellar populations present in
a galaxy can be done from a quick look to a good CMD. The presence of stars
in characteristic evolutionary phases indicates that star formation took
place in the system in one or another epoch of its history. For example, the
presence of RR-Lyrae stars is indicative of an old, low metallicity stellar
population; a substantial amount of red-giant branch (RGB) stars is associated
to an intermediate-age to old star formation activity; a well developed
red-tail of asymptotic giant branch (AGB) stars shows that intermediate-age
to young stars with relatively high metallicity are present in the system,
and even a few blue, bright stars as well as HII regions are evidences of a
very recent star formation activity. 

A higher degree of sophistication is provided by isochrone fitting to
significant features of the CMD. Indeed, this method is simple and powerful
enough to determine age and metallicity of simple stellar populations as the
ones present in star clusters. However, actually deciphering the information
contained in a complex CMD and deriving a quantitative, accurate SFH is
complicated and requires some relatively sophisticated technique. Although
other approaches are possible (see below), the most extended and probably
most powerful technique is the one based on synthetic CMD analysis. The
standard procedure involves three main ingredients: (i) a CMD ideally reaching the oldest main-sequence turn-offs; (ii) one (or several) synthetic CMDs, computed assuming a set
of input physical parameters, and (iii) a method to derive the SFH from the
comparison of the star distributions in observational and synthetic CMDs. 

In Aparicio \& Gallart (2004) we presented IAC-star, a code for synthetic CMD
computation. In short, the algorithm is intended to be as general as possible
and allows a variety of inputs for the initial mass function (IMF), star
formation rate, metallicity law and binariety. Stars with age and metallicity
following a continuous distribution are computed through interpolation in a
stellar evolution library, providing synthetic CMDs with smooth, realistic
stellar distributions.

In this paper we present an algorithm corresponding to ingredient (iii);
i.e., designed for deriving the SFH from the comparison of observed and
synthetic CMDs. Several approaches have been used in the past. Indeed,
interest in the process of formation of stellar populations is not new. The
early works by Baade (1944) can be considered a starting point. However, the
amount of information and results has been largely increased in the last few
decades and it is only since relatively recently that we have properly been
able to speak about quantitative determinations of complex SFHs. Tosi et al.
(1991) used a method based on the comparison of luminosity functions of
observed and synthetic CMDs. They were the first to sketch SFHs of nearby
galaxies using synthetic CMDs. Bertelli et al. (1992) introduced the
R-method, the first to make use of the global morphology and number counts of
the distribution of stars in synthetic and observed CMDs to derive the SFH of
a galaxy (the LMC). Gallart et al. (1996) used an extended, more complete
version of the R-method to study the SFH of the Local Group galaxy NGC 6822. Tolstoy \& Saha (1996) used maximum likelihood to find out the synthetic CMD, of a set of them, best
reproducing the observed one. Vergely et al. (2002) presented an inverse method to interpret the CMD in terms of the SFH.

Aparicio, Gallart, \& Bertelli (1997)
introduced a new approach in which the SFH is derived by linear combination
of several simple populations. This method allows a SFH 
derivation free from initial assumptions about it and has the important advantage of requiring only a single synthetic CMD. The same idea was independently
introduced by Dolphin (1997) and has also been applied by Holtzman et al.
(1999), who, for the first time, derived the star formation rate (SFR) as a function of time and chemical enrichment law (CEL) simultaneously, making no assumptions about the CEL morphology; by Olsen (1999), by Harris \& Zaritsky (2001) and by Rizzi et al. (2003) and has been used in several works of the former authors (e.g. Gallart et al. 1999; Aparicio, Tikhonov, \& Karachentsev 2000; Holtzman, Smith, \& Grillmair 2000; Dolphin 2000a,b; Aparicio, Carrera, \& Mart\'{\i}nez-Delgado 2001; Wyder 2001; Sabbi et al. 2007; Chiosi, \& Vallenari 2007; see Aparicio 2002 for a review). 

The former methods do in general include some kind of parametrization of the
CMD. It must also be mentioned in this short historical overview the
contribution by Hern\'andez, Valls-Gabaud, \& Gilmore (1999) as an
alternative, non-parametric method for the derivation of SFH of galaxies
which make direct use of the information contained in a stellar evolution
library and it is not based in synthetic CMDs.

The code we present here is based in the same principle as that used by
Aparicio et al. (1997) but makes use of a genetic algorithm ({\it pikaia};
Charbonneau 1995) for the solution convergence. Applied in the most
general way it derives the SFR as a function of both time and metallicity or, from a different point of view, the SFR and the CEL as a function of time of a system from the comparison of
its CMD star distribution with the star distribution in a single template
synthetic CMD.

The code is made available for free use. It can be downloaded from the internet
site {\tt http://iac-star.iac.es/iac-pop} with the only requirement of referencing this paper and crediting as indicated in that site. It is worth mentioning that the code is not restricted to the SFH solution only, but to any problem that can be parametrized in a similar way. The present paper should be considered as the
reference for the code. This paper is a complement of our previous one
presenting IAC-star, the code for synthetic CMD computation (Aparicio \&
Gallart 2004).

This paper is organized as follows. In \S\ref{solvsfh}, the method is presented. In \S\ref{test}, \S\ref{error}, \S\ref{bursts}, and \S\ref{general}, the IAC-pop code self-consistency, error sources and reliability of solutions are discussed under different input assumptions. In \S\ref{run} a cook-book for IAC-pop execution is given. Finally, in \S\ref{conclusions} some final remarks are done and the main conclusions are summarized.

\section{IAC-pop: a method to solve the Star Formation History \label{solvsfh}}

\subsection{Basic definitions}

The SFH is composed by several pieces of information. The rate at which stars form as a function of time (SFR) and the metallicity distribution of those stars, also a function of time (CEL), are the most important characteristics. The initial mass function and the binariety of stars are also related with the SFH (see Aparicio 2002). For simplicity we will adopt here the following approach: considering that time and metallicity are the most important variables in the problem, we define the SFH as a function $\psi(t,z)$ such that $\psi(t,z){\rm d}t{\rm d}z$ is the number of stars formed at time $t'$ in the interval $t<t'\leq t+{\rm d}t$ and with metallicity $z'$ in the interval $z<z'\leq z+{\rm d}z$, per unit time and metallicity. $\psi(t,z)$ is a distribution function and can be identified with the usual SFR, but as a function both of time and metallicity. 

There are several other functions and parameters related to
the SFH, that we will consider here as auxiliary. The aforementioned IMF, 
$\phi(m)$ and a function accounting for the frequency and relative
mass distribution of binary stars, $\beta(f,q)$ are the main ones. The
solution found for the SFH depends on the assumptions made for $\phi(m)$ and
$\beta(f,q)$. In the most general case, they should be free and solved
together with the SFH. In practice, the amount of available 
information may not be sufficient to attempt such an assumption free solution, but, in any case, several choices of both
$\phi(t)$ and $\beta(f,q)$ should be tried.

Other parameters affecting the solution of $\psi(t,z)$ are distance and
reddening, including differential reddening. But the strongest limitation on
the observational information is produced by the {\it observational
effects}. These include all the factors affecting and distorting the
observational material, namely the signal-to-noise limitations, the defects
of the detector and the crowding and blending between stars. The consequences
are loss of stars, changes in measured stellar colors and magnitudes, 
and external errors larger and more difficult to control than internal
ones. Comparison of Figs. \ref{f02_dcm_noerr} and \ref{f12_dcm_err} show the distortion introduced by observational effects (see below).
It must be kept in mind that IAC-pop does not account for the observational effects, that must be considered
and simulated in advance by the user into the synthetic CMDs.

In the following, we will concentrate in the determination of $\psi(t,z)$ on the understanding that
all the remaining functions and parameters are externally checked and that
proper assumptions are made for them.

\subsection{The IAC-pop methodology in short \label{parametrizing}}

The procedure used by IAC-pop is based on the one introduced independently
by Aparicio et al. (1997) and Dolphin (1997). Its fundamentals have been
adopted and extended by several groups since then (see \S\ref{intr}). The method is based on the following steps:

\begin{enumerate}
  
\item The synthetic CMD computation code IAC-star or any other code intended for the same purpose is used to generate a
  single global synthetic stellar population with a large number of stars with ages and
  metallicities following some convenient distribution over the full interval of variation of
  $\psi(t,z)$ in time and metallicity. The simplest case is using a constant distribution, but other approaches would be convenient in order to better sampling some age or metallicity intervals. Observational effects (crowding, blending, external errors, etc) should be
  simulated in the synthetic CMD.
  
\item The former synthetic stars are distributed in an array of partial or simple models (see Fig. \ref{f03_sfhsintetica} below). Each contains the stars within small intervals of age and metallicity. They constitute a set of $n\times m$ models with no star in common between any two of them. Arbitrary stellar populations can be obtained by linear combinations with non negative coefficients of the simple models of this set. These properties are similar to those of a base of a vectorial space. However, the simple model set can not be defined as such because of their statistical nature and because of the fact that coefficients can not be negative. 

\item A set of boxes is defined in the CMD. In practice, several approaches are possible, including uniform and {\it a la carte} grids. An uniform grid is more objective and less dependent on human criteria. An
  {\it a la carte} grid takes advantage of our knowledge of stellar evolution
  and allows different sampling of well and poorly known stellar evolution
  phases. The approach we use here is a bit more sophisticated and takes advantage of the strengths of the two aforementioned approaches. Several regions, that we will call {\it bundles}, are defined in the CMD. Each of them are sampled by an uniform grid, but the grid bin size can be different from one bundle to another (see Fig. \ref{f04_dcms_noerr} below).
  
\item An array, $M_i^j$, containing the number of stars from partial model
  $i$ populating the CMD box $j$ is computed. The same operation is made in the
  observational CMD, producing a vector, $O^j$, containing the number of
  observed stars in box $j$. This step defines the parametrization of the
  CMD. 
  
\item With the former information, the distribution of stars in the CMD
  boxes can be calculated for any model SFH as a linear combination of the $M_i^j$ values:

\begin{equation}
M^j=A\sum_i\alpha_iM_i^j
\end{equation}

It should be noted that $\alpha_i\geq 0$. $A$ is a scaling constant.

\item The SFH best matching the distribution, $O^j$, of the observational CMD
  can be found using a merit function. In particular Mighell's $\chi^2_\gamma$ (Mighell 1999) is used:

\begin{equation}
\chi^2_\gamma=\sum_j\frac{(O^j+min(O^j,1)-M^j)^2}{O^j+1}
\end{equation}

We will use $\chi^2_\nu=\chi^2_\gamma/\nu$, where $\nu$ is the number of freedom degrees. In our case $\nu=k-(n\times m)$, where $k$ is the number of boxes defined in the CMD.

Minimization of $\chi^2_\nu$ provides the best solution as a set of $\alpha_i$ values, as well as a test on whether it is good enough. IAC-pop makes use of a genetic algorithm for an efficient searching of the $\chi^2_\nu$ minimum. Such a procedure is required because of the large number of the problem dimensions ($n\times m$).

\item The solution SFH can be written as:

\begin{equation}
\psi(t,z)=A\sum_i\alpha_i\psi_i
\end{equation}

where $\psi_i$ refers to partial model $i$, with $i$ taking values from 1 to
$n\times m$, and $A$ is a scaling constant.

\end{enumerate}

\section{Running IAC-pop}\label{test}

\begin{figure}[ht!]
\plotone{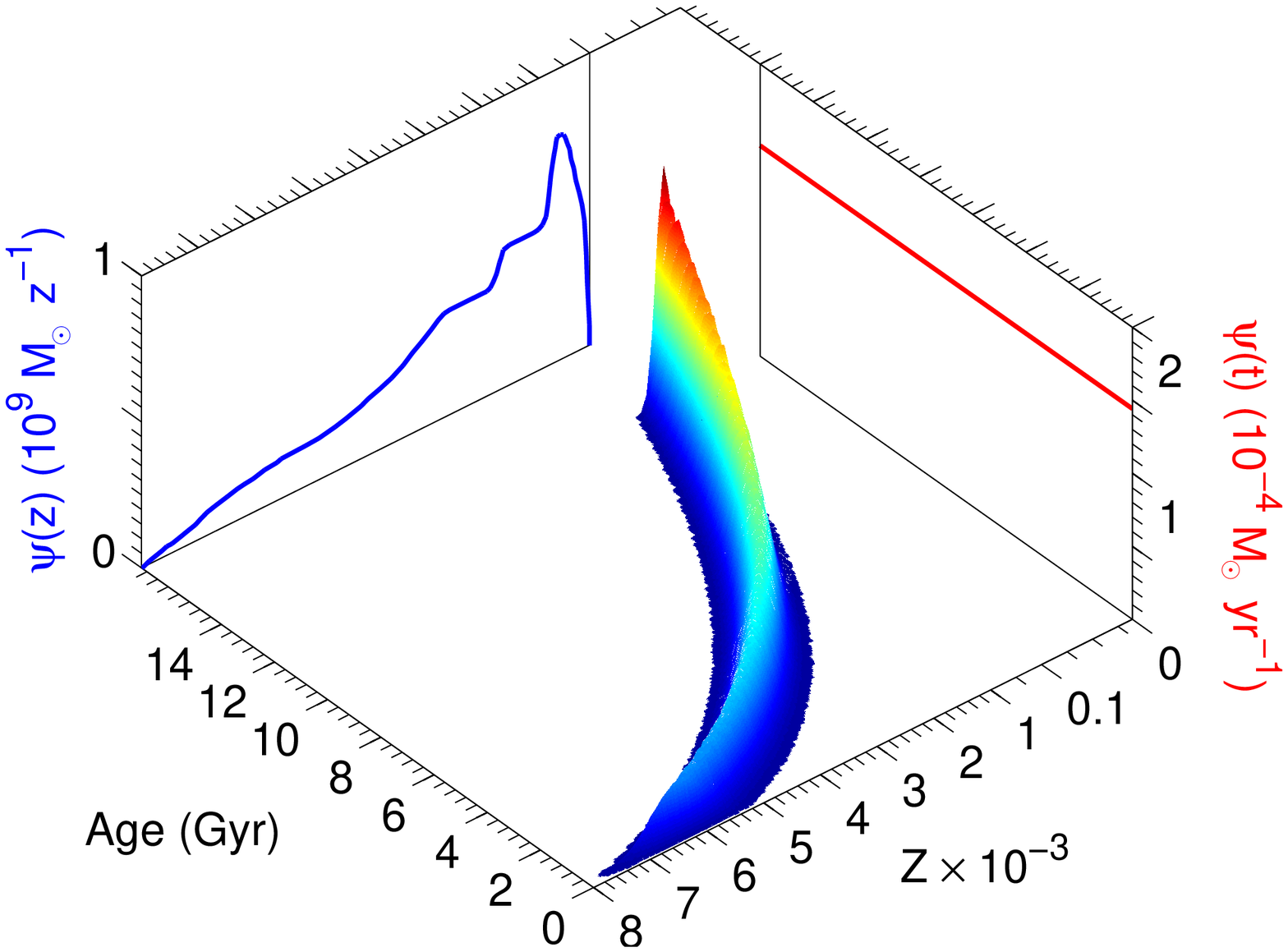}%
\figcaption[]{The SFH $\psi(t,z)$ of the mock population. Age and metallicity are given in the horizontal axis. The volume below the curved surface and over the age-metallicity plane gives the mass that has been ever transformed into stars within the considered age-metallicity interval. The mono-dimensional $\psi(t)$ and $\psi(z)$ are shown on the $\psi$-age plane (in red in the paper electronic version) and on the $\psi$-metallicity plane (in blue in the paper electronic version) respectively.
\label{f01_sfhmodelo}}
\end{figure}

We have made several tests to check IAC-pop efficiency, self-consistency, solution stability and whether it can deal with real astrophysical data. To this purpose, a synthetic stellar population has been generated with IAC-star from some arbitrary SFH. We will call this the {\it mock} population and denote as mCMD the corresponding CMD. The mock population is solved as if it were real data and the solution is compared with its input SFH. The IAC-star input parameters used to compute the mock population were as follows. The Teramo-BaSTI stellar evolution library (see Pietrinferni et al. 2004) and the Castelli \& Kurucz (2003) bolometric correction library were used. The number of stars in the mCMD was $10^5$. The star formation ranges from 14 Gyr ago to date with a constant SFR, $\psi(t)$, for that period. The metallicity increases with time, with initial and final metallicities $z_0 = 0.0001$ and $z_f=0.008$ and some metallicity dispersion at each time (see Fig. \ref{f01_sfhmodelo}). Finally, no binary stars were considered and the IMF by Kroupa, Tout \& Gilmore (1993) was used. The integral of $\psi(t,z)$ (i.e. the total mass ever transformed into stars) for this system is $\Psi_T=2.02\times 10^6 M_{\sun}$.

\begin{figure}[ht!]
\plotone{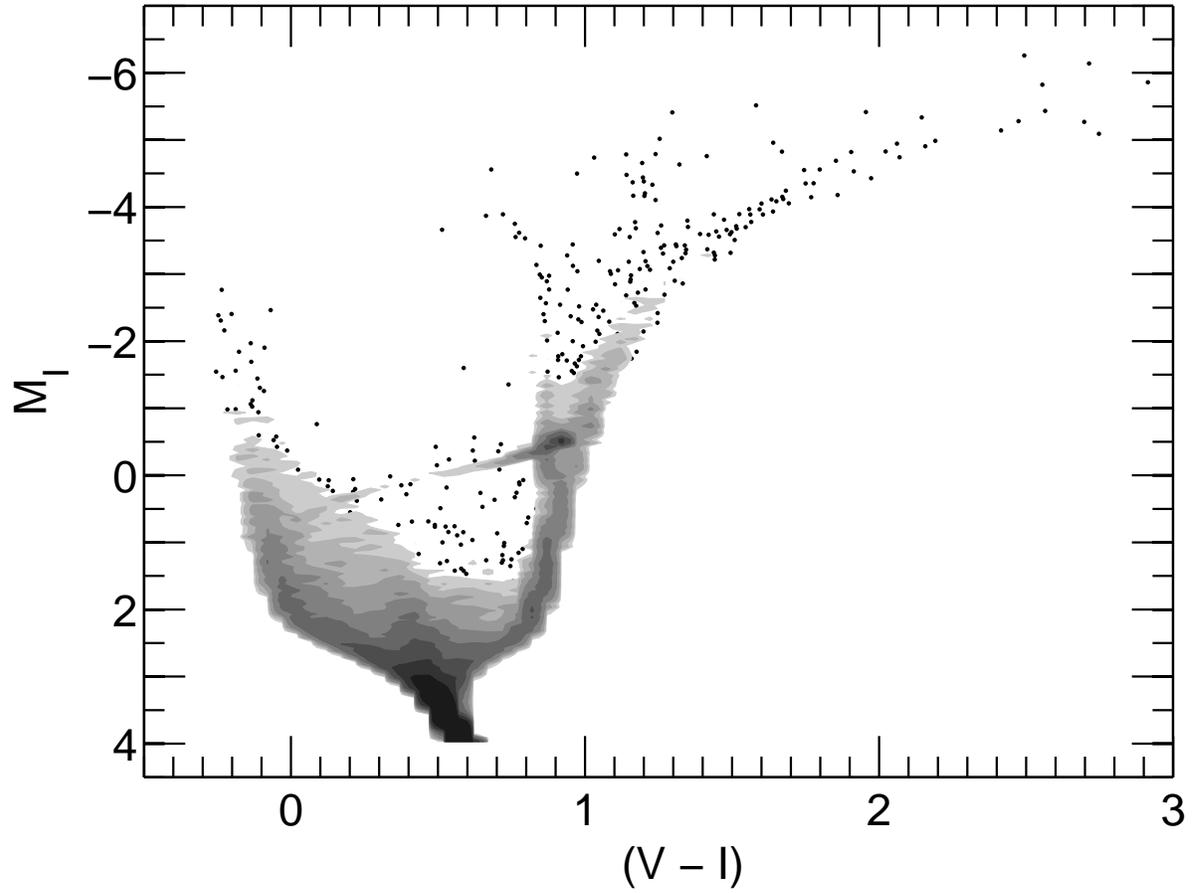}%
\figcaption[]{The CMD of the mock population (mCMD). Grey levels show the density of stars. A factor of two in density exists between each two successive gray levels. Single dots are shown where the density is less than 8 stars per $(0.1)^2$ magnitudes. Grey levels start respectively at 8, 16, 32, 64, 128, 256, 512, 1024 and 2048 stars per $(0.1)^2$ mag interval. 
\label{f02_dcm_noerr}}
\end{figure}

The SFH $\psi(t,z)$ of the mock population is shown in Fig. \ref{f01_sfhmodelo}. The volume below the curved surface and over the age-metallicity plane gives the mass that has been ever transformed into stars within any considered age-metallicity interval. The SFR as a function of time only, $\psi(t)$, and of metallicity only, $\psi(z)$, are also shown. The mCMD is shown in Fig. \ref{f02_dcm_noerr}.

\begin{figure}[ht!]
\plotone{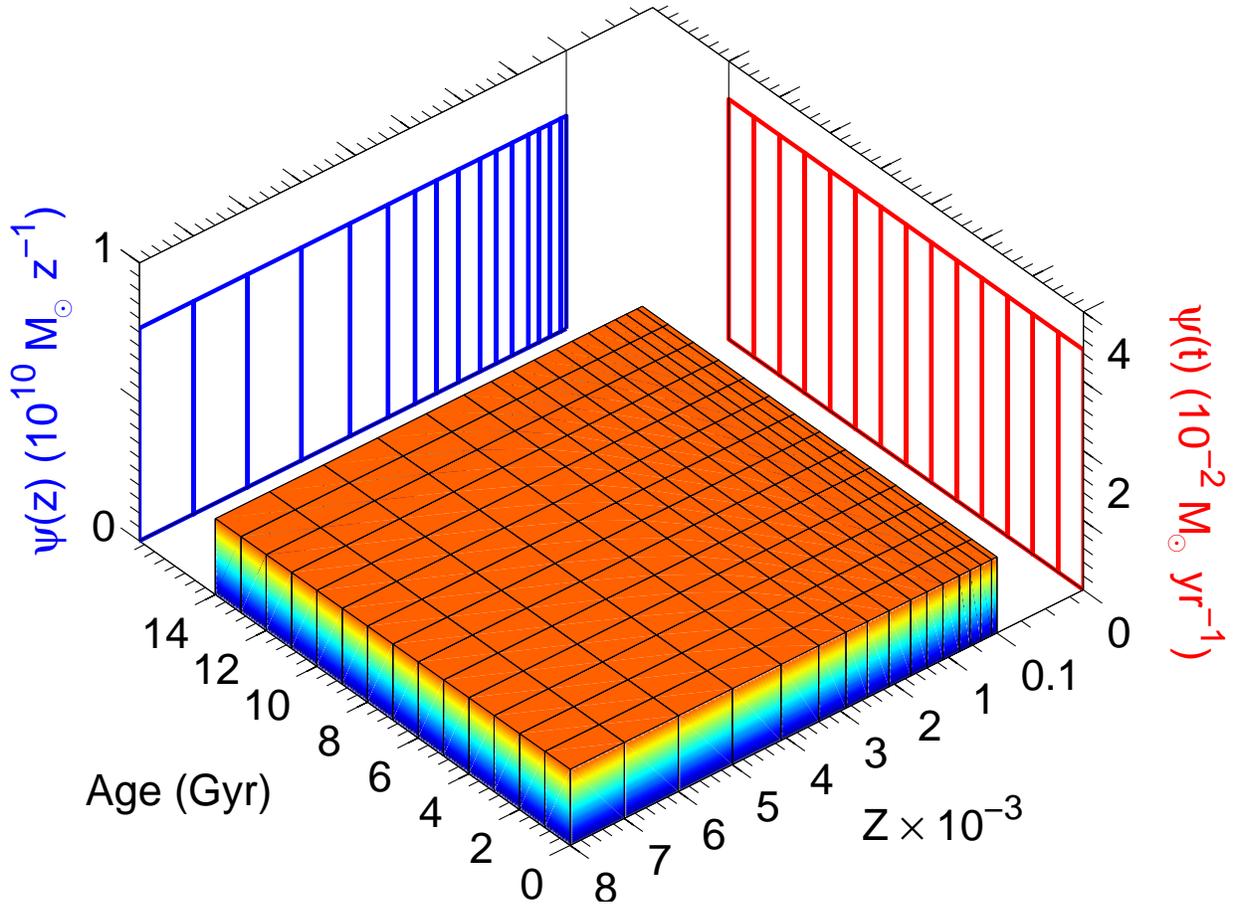}%
\figcaption[]{Star Formation History of the global synthetic population associated to sCMD. The caption is the same as in Fig. \ref{f01_sfhmodelo}. Bars shown the simple populations in which the global one has been divided for the analysis of the problem. See text for details. 
\label{f03_sfhsintetica}}
\end{figure}

\begin{figure}[ht!]
\plotone{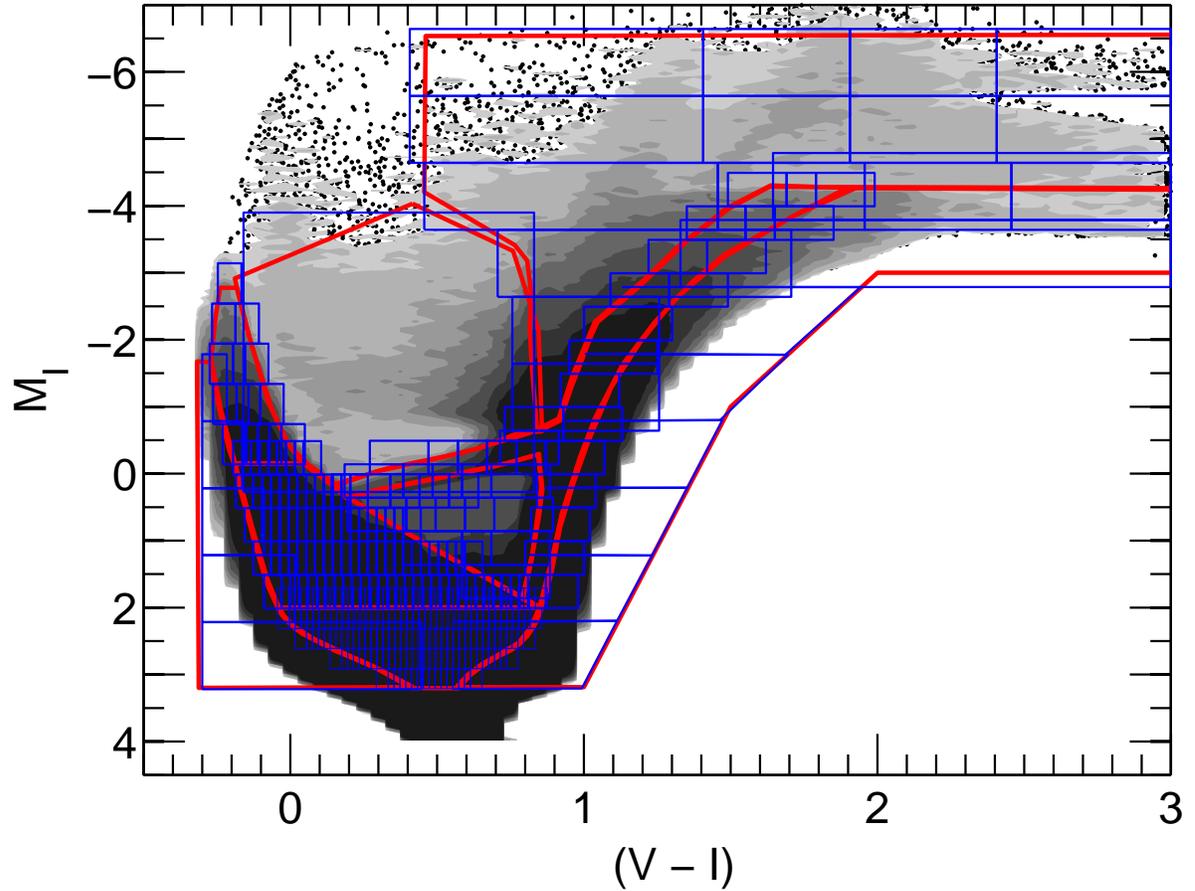}%
\figcaption[]{The CMD of the global synthetic stellar population (sCMD). Grey levels show the density of stars as in Fig. \ref{f02_dcm_noerr}. Bundles (in red in the electronic version) and one of the used box distributions (in blue in the electronic version) are shown.
\label{f04_dcms_noerr}}
\end{figure}

According to \S\ref{parametrizing}, item 1, a global synthetic population has been computed as the starting point for the solution searching. We will call its CMD, sCMD. Binary stars, IMF and stellar evolution and bolometric correction libraries used are the same as for the mock population. A constant SFR, $\psi(t,z)$, was used for the full age (from 0 to 14 Gyr) and metallicity (from 0.0001 to 0.008) ranges. The sCMD contains $3\times 10^7$ stars. The SFH of this population is shown in Fig. \ref{f03_sfhsintetica}. Age and metallicity have been divided into 14 and 16 intervals respectively. These intervals define one of the simple population sets which have been used for the solution searching, as mentioned in \S\ref{parametrizing}, item 2. The sCMD is shown in Fig. \ref{f04_dcms_noerr}.

\placefigure{f01_sfhmodelo}
\placefigure{f02_dcm_noerr}
\placefigure{f03_sfhsintetica}
\placefigure{f04_dcms_noerr}

As mentioned in \S\ref{parametrizing}, item 3, several bundles have been defined onto the mCMD and the sCMD. Each one has been sampled by grids of different box sizes, which depend on the CMD region or bundle. Grids are thinner in regions where age and metallicity resolution is better, like the main sequence (MS), and coarser in regions in which the distribution of stars strongly depends on poorly known parameters, like the red giant branch (RGB) or the horizontal branch (HB). The number of boxes in a grid ranges from one to several hundreds depending on the CMD region. Figure \ref{f04_dcms_noerr} shows the bundles and one of the used box distributions, overplotted on the sCMD.

The particular box distribution used and the division of the global synthetic population in several simple populations as a function of which the solution is searched may introduce some bias and binning effects on such solution. In order to minimize them, several sets of grids and simple populations have been used. Each one is defined by shifting the set with respect to the initial one. In particular, in our case, bundles were fixed, and a total of 3 grid sets and 40 partial model sets were used and a solution computed for each one. The average of all them is adopted as the final solution. Our many tests have disclosed that this procedure provides very stable solutions and the best way to estimate errors (see \S\ref{error}).

A set of routines called MinnIAC is used to control the full computation process. The MinnIAC input consists in a set of parameters that define the (i) the bundles to be used in the CMDs; (ii) the grid size in each bundle; (iii) the number of different grids to be used and the shifts to be applied to define them; (iv) the division of the global synthetic population into simple populations; (v) the number of different input sets of simple populations and the shifts to be applied to define them. For each set, MinnIAC counts stars in the grids of the observational and simple populations CMDs and generates an input parameter file for IAC-pop. Once all the solutions have been obtained, MinnIAC averages them and computes errors for each age and metallicity interval as the root mean square of the solutions. MinnIAC will be provided under request but should be used at user's risk, since no support can be provided by the authors at this moment. 

\begin{figure}[ht!]
\plotone{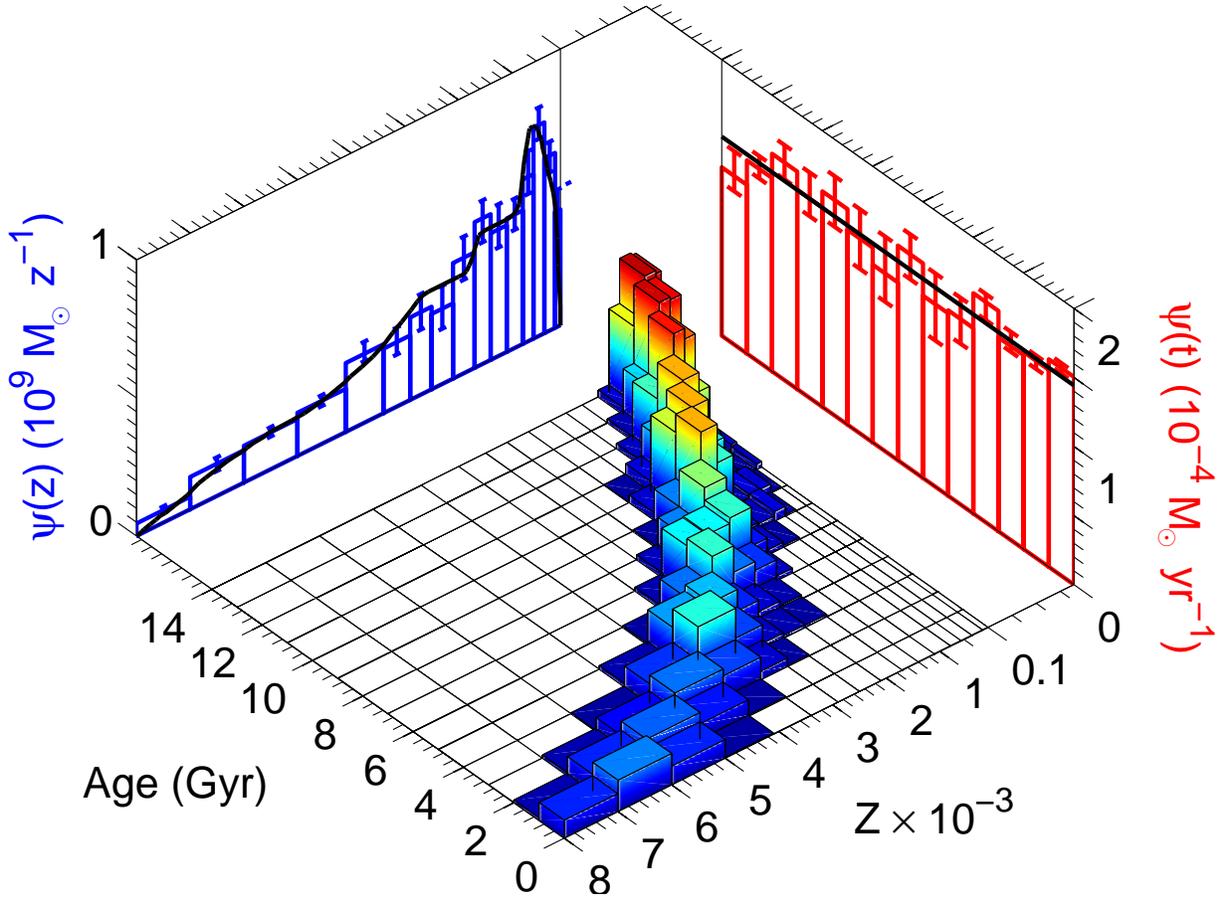}%
\figcaption[]{Solution of $\psi(t,z)$ obtained for the mock population. No observational errors have been simulated. It is the average of 120 single solutions found using the {\it several solutions} procedure described in text.  The caption is the same as in Fig. \ref{f01_sfhmodelo}. Projection of the solution onto the $\psi(t)$-age and $\psi(z)$-$Z$ planes, obtained as integration of $\psi(t,z)$ over metallicity and time, respectively, are also shown. The mono-dimensional input $\psi(t)$ and $\psi(z)$ are shown by solid lines (see Fig. \ref{f01_sfhmodelo}). Error bars have been obtained applying the {\it several solutions} procedure.
\label{f05_nocrow}}
\end{figure}

\placefigure{f05_nocrow}
\placefigure{f06_sol_nocrow4}
\placefigure{f07_genecrep}

\begin{figure}[ht!]
\plotone{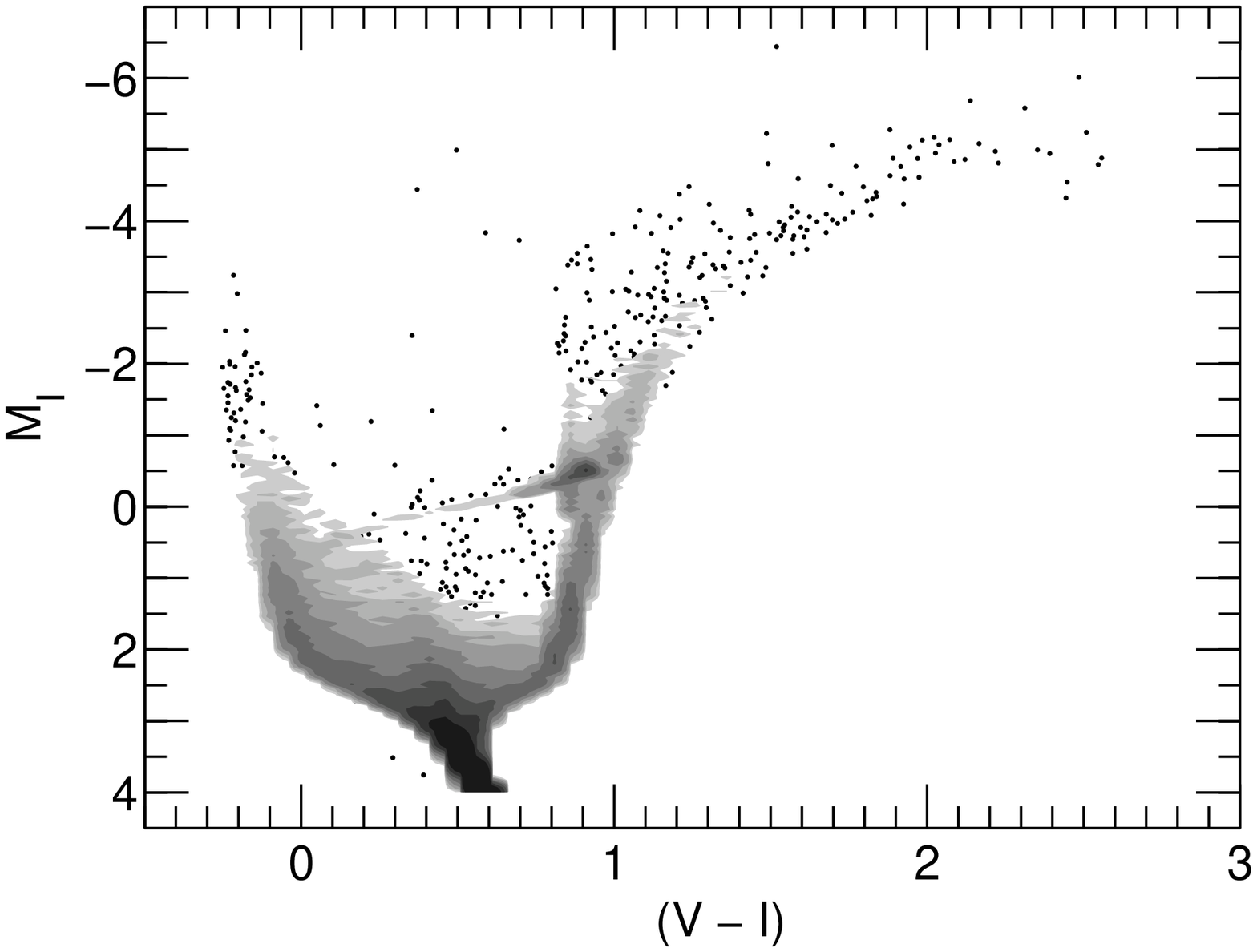}%
\figcaption[]{The CMD corresponding to the SFH shown in Fig. \ref{f05_nocrow}. To be compared with the mCMD shown in Fig. \ref{f02_dcm_noerr}
\label{f06_sol_nocrow}}
\end{figure}

\begin{figure}[ht!]
\plotone{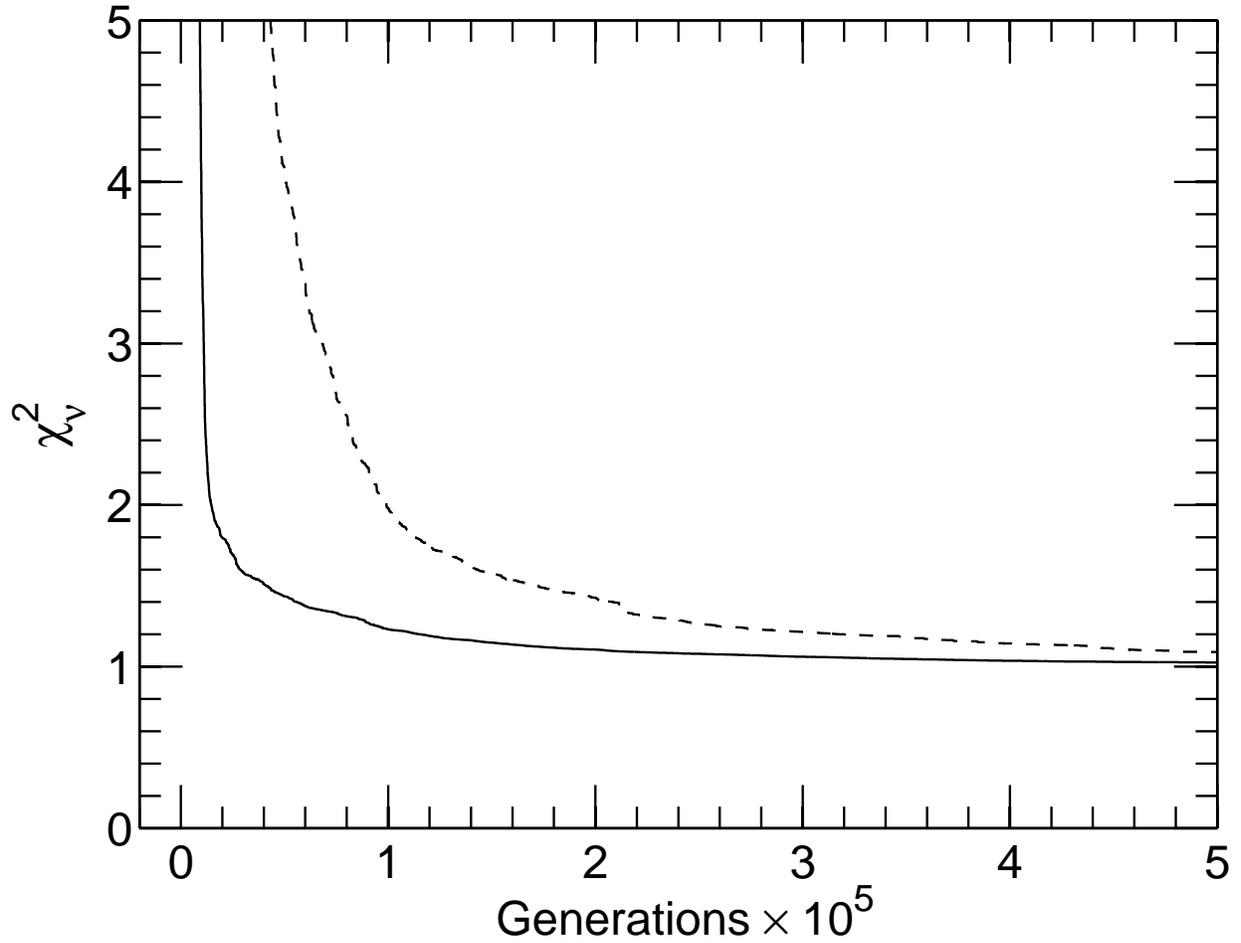}%
\figcaption[]{$\chi^2_\nu$ as a function of the number of IAC-pop genetic generations. The latter is proportional to the computing time. Dashed and solid lines show $\chi^2_\nu$ for two different selections of input values of genes, size of the population and mutation rate, which are internal parameters of the genetic algorithm.
\label{f07_genecrep}}
\end{figure}

Figure \ref{f05_nocrow} shows the $\psi(t,z)$ solution for the mock population given in Fig. \ref{f01_sfhmodelo}. Comparison between input and solution is simpler for the monodimensional $\psi(t)$ and $\psi(z)$ functions, also shown in Fig. \ref{f05_nocrow}. Agreement is good, being the differences between input and solution within the error bars (see below the discussion of error computation). As a further visual test, Fig.\ref{f06_sol_nocrow} shows the CMD corresponding to the solution shown in Fig. \ref{f05_nocrow}. It is to be compared with the input mCMD shown in Fig. \ref {f02_dcm_noerr}.  

As mentioned above, IAC-pop uses a genetic algorithm to look for the set of positive or null $\alpha_i$ parameters minimizing the $\chi^2_\nu$ function given in \S\ref{parametrizing}, item 6. The convergence velocity depends on the choice of input parameters for the genetic code (see Charboneau 1995). Figure \ref{f07_genecrep} shows the evolution of $\chi^2_\nu$ as a function of generation for our test case. Dashed and solid lines show the $\chi^2_\nu$ for two different selections of input values of genes, number of individuals and mutation rate of IAC-pop. Both inputs get almost the same $\chi^2_\nu$ if the number of generations is large enough. In other words, changes in the genetic input values do not affect the final solution if the number of generations is large, but computation time, which is proportional to the number of generations, may be large if the parameter choice is not good.

\section{Error sources}\label{error}

Providing reliable error estimates is fundamental for the solution meaningfulness. In principle, we can identify three kinds of error sources in the problem of retrieving the SFH from a CMD. The first one is that of internal errors, inherent to the numerical problem or related to the parametrization, i.e. the way in which bundles, grids and also partial models are selected. The second one is that associated to observational effects (crowding, blending, etc). They blur the CMD, distorting the information provided by it. Observational effects must be simulated in the global synthetic CMD before parametrizing it. Also, any estimate of errors of the first kind will include the observational effects. Finally, an additional uncertainty source is related to our limited knowledge of stellar evolution theory. It is difficult to test this, but using different stellar evolution libraries will produce somewhat different solutions for the same input observational data. Differences between these solutions provide a representation of this kind of uncertainty. In the following we will discuss each of the former kinds of error sources.

\subsection{Errors of the solution}\label{internos}

Calculating the internal error of the best solution for the SFH is not straightforward. The $\chi^2$ test allows computing $1\sigma$ errors of the solution parameters (see Bevington \& Robinson 2003 and Arndt \& MacGregor 1966). But this works only for Gaussian random variables. The fact that the SFH problem can depart significantly from Gaussian makes this approach unuseful for our purposes, as we have checked in several examples not shown here for brevity. 

Errors related to the data sampling can be calculated assuming that the number of stars in each CMD box behave according to a Poisson statistics. We will call this the {\it Poisson statistics} criterion. Errors are computed as follows. Once the best solution has been found, the input observational data, i.e. the number of stars in each grid box, are randomly modified according to a Poisson statistics. The best solution for the new data set is computed. This procedure is repeated several times, providing a set of solutions, the average of which is expected to be similar to the best original one. Also, if $n_r$ such solutions have been obtained and solution $r$ is given by the set $\alpha_{r,i}$, then 

\begin{equation}
\sigma_{\alpha_i}=\left(\frac{\sum_{r}(\alpha_{r,i}-\alpha_i)^2}{n_r-1}\right)^{1/2}
\end{equation}

\noindent can be used as an estimate of the errors affecting $\alpha_i$ arising from data sampling. These errors are provided by IAC-pop. Upward and downward error bars can be obtained using only the subsets $\alpha_{r,i}\geq\alpha_i$ and $\alpha_{r,i}\leq\alpha_i$, respectively and are also provided by IAC-pop.

The former criterion does not include all error sources and it will likely produce error underestimates. In particular, it does not take into account the effects introduced by the CMD parametrization through bundles and grids nor those related to the division of the global synthetic CMD into partial models. A simple way to evaluate these errors is using the dispersion of the 120 solutions computed and explained in \S\ref{test}. This is the procedure used in this paper. We will see below that it is a good estimate of all the internal error sources. For brevity, we will call this the {\it several solutions} criterion. The error bars shown in all the figures have been calculated in this way. It is time consuming, but we think it is necessary if a stable solution and a realistic estimate of errors is sought. The routine MinnIAC delivers the errors computed in this way (see \S\ref{test}).

\begin{figure}[ht!]
\plotone{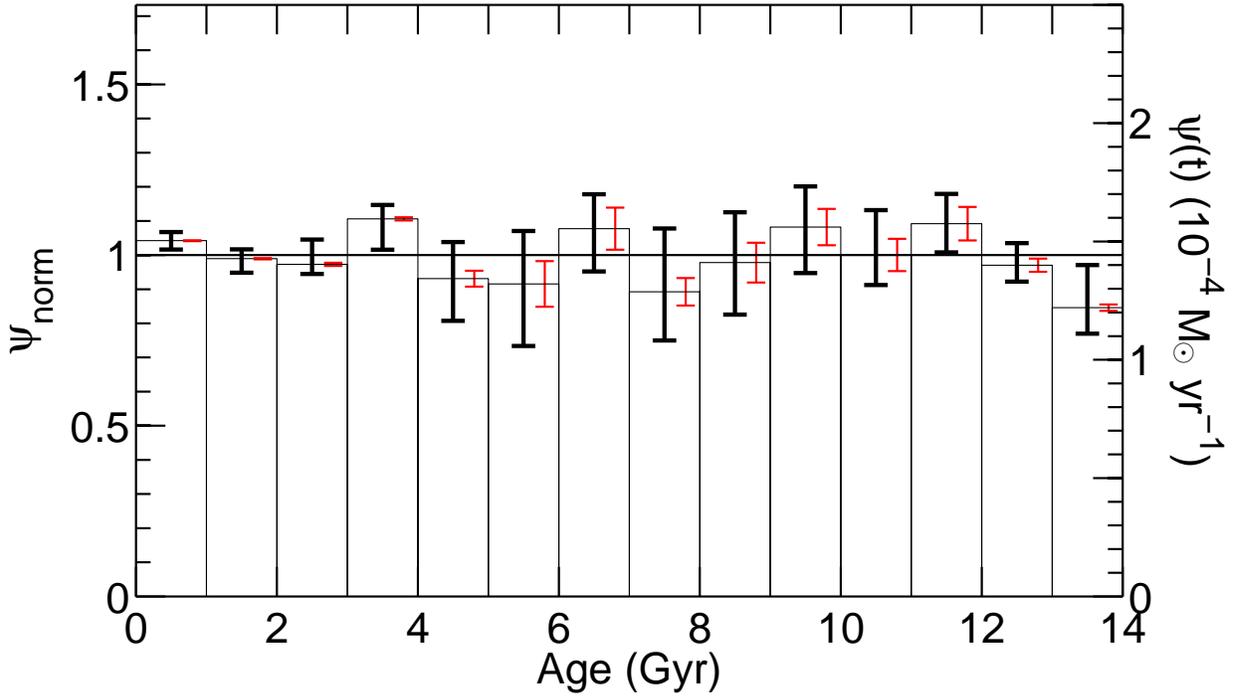}%
\figcaption[]{The projection onto the $\psi(t)$-age of the solution for the SFH given in Fig. \ref{f05_nocrow} is shown. The horizontal line corresponds to the input $\psi(t)$. Thick error bars have been obtained applying the {\it several solutions} procedure described in text. In this case, 24 solutions have been used. Thin (red in the electronic version of the paper) error bars result from the {\it Poisson statistics} formal error computation. See text for details.
\label{f08_minniac1}}
\end{figure}

\begin{figure}[ht!]
\plotone{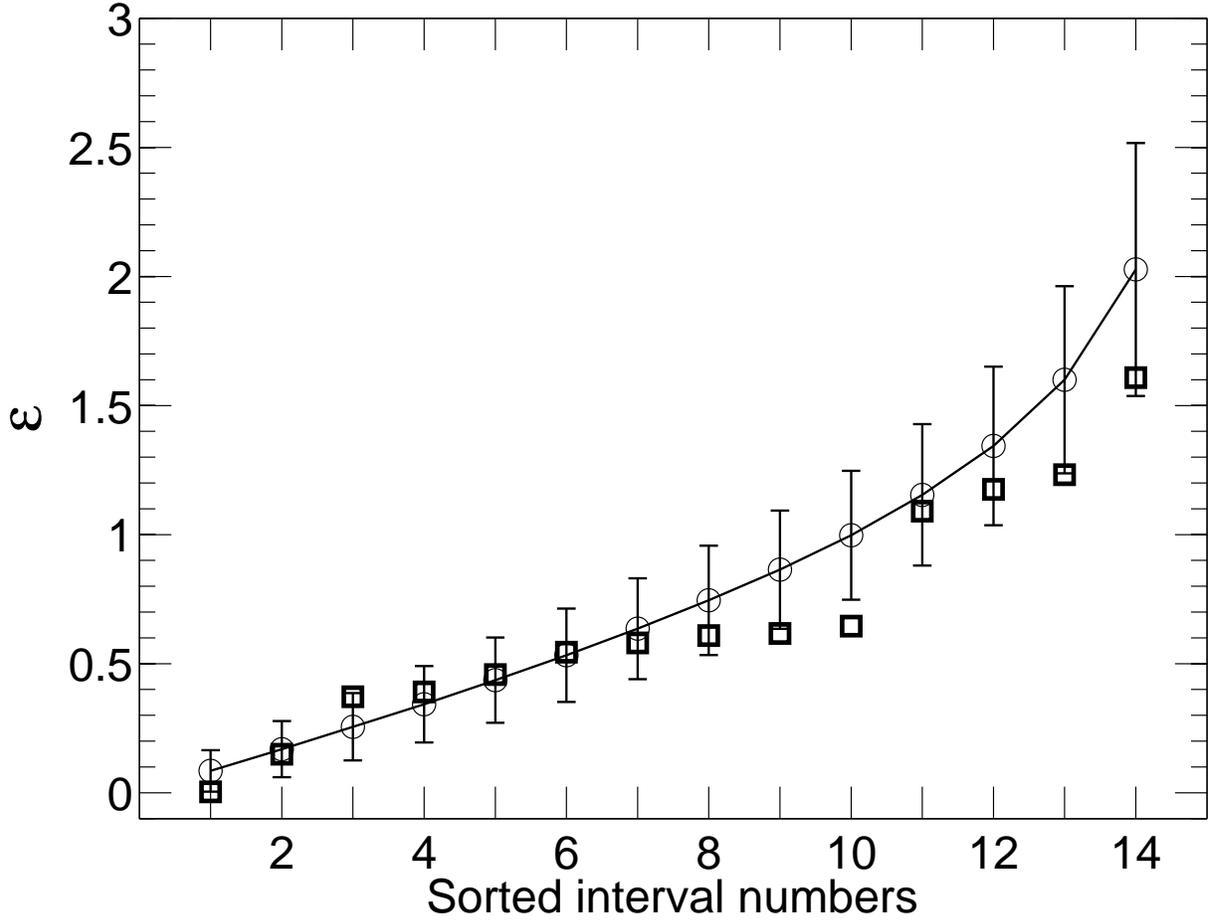}%
\figcaption[]{The $\epsilon$ parameter is represented as a function of age interval order number, after sorting by increasing values of $\epsilon$. The parameter is defined as $\epsilon_i=|\psi^{\rm out}_i(t)-\psi^{\rm in}_i(t)|/\sigma_i$ where $i$ stands for the age interval and $\psi^{\rm out}_i(t)$ and $\psi^{\rm in}_i(t)$ refers to the output (or solution) and input values of $\psi(t)$, respectively. In other words, $\epsilon$ parameters are the absolute values of the differences between solution and input measured in units of the corresponding $\sigma$ values. Thick squares show the values of the solution given in Figs. \ref{f05_nocrow} and \ref{f08_minniac1}. Thin circles and the line connecting them show the behavior of a normalized Gaussian random variable. $10^5$ experiments in which 14 values are randomly selected have been done to obtain this. Circles show the averages of the $10^5$ experiments and error bars give their rms dispersions.
\label{f09_minniac2}}
\end{figure}

\placefigure{f08_minniac1}
\placefigure{f09_minniac2}

Figure \ref{f08_minniac1} shows the input and solution $\psi(t)$ together with the error obtained by the {\it several solutions} (thick, black error bars) and the {\it Poisson statistics} (thin, red error bars) criteria. That the latter are underestimates of total errors should be clear from the fact that in 12 out of 14 intervals, differences between solution and input are larger than the error bars and, in several cases, much larger than three times that. However, this does not seem to be the case for the errors computed by the {\it several solutions} criterion. To test if the errors obtained in this way are similar to what should be expected in a Gaussian case the following test has been done. For each age interval, the parameter $\epsilon_i=|\psi^{\rm out}_i(t)-\psi^{\rm in}_i(t)|/\sigma_i$ has been computed, where $i$ stands for the age interval, $\sigma_i$ is the rms of the solutions and $\psi^{\rm out}_i(t)$ and $\psi^{\rm in}_i(t)$ refer to the output (or solution) and input values of $\psi(t)$, respectively. In other words, $\epsilon$ parameters are the absolute values of the differences between solution and input measured in units of the corresponding $\sigma$ values. The $\epsilon$ parameters have been sorted and plotted in Fig. \ref{f09_minniac2} together with the values of a standard Gaussian random variable. For the latter, $10^5$ experiments have been done. For each one, 14 values have been randomly given to the random variable and then sorted from smaller to larger. The values represented in Fig. \ref{f09_minniac2} (open circles) are the averages for the $10^5$ experiments while the errors bars show the corresponding rms dispersions. It can be seen that a reasonable agreement exists between our results and the Gaussian case, indicating that the {\it several solutions} approach produces reliable estimates of total internal errors. 

\subsection{Including observational effects}\label{crowding}

Observational effects include limited signal-to-noise, incompleteness, source blending, image read-out noise as well as not fully removed artifacts in the data reduction process (flat-field correction, etc). All together, their effects on the CMD are the loss of stars and dispersion and shift of points, all depending on magnitude and color. 

\begin{figure}[ht!]
\plotone{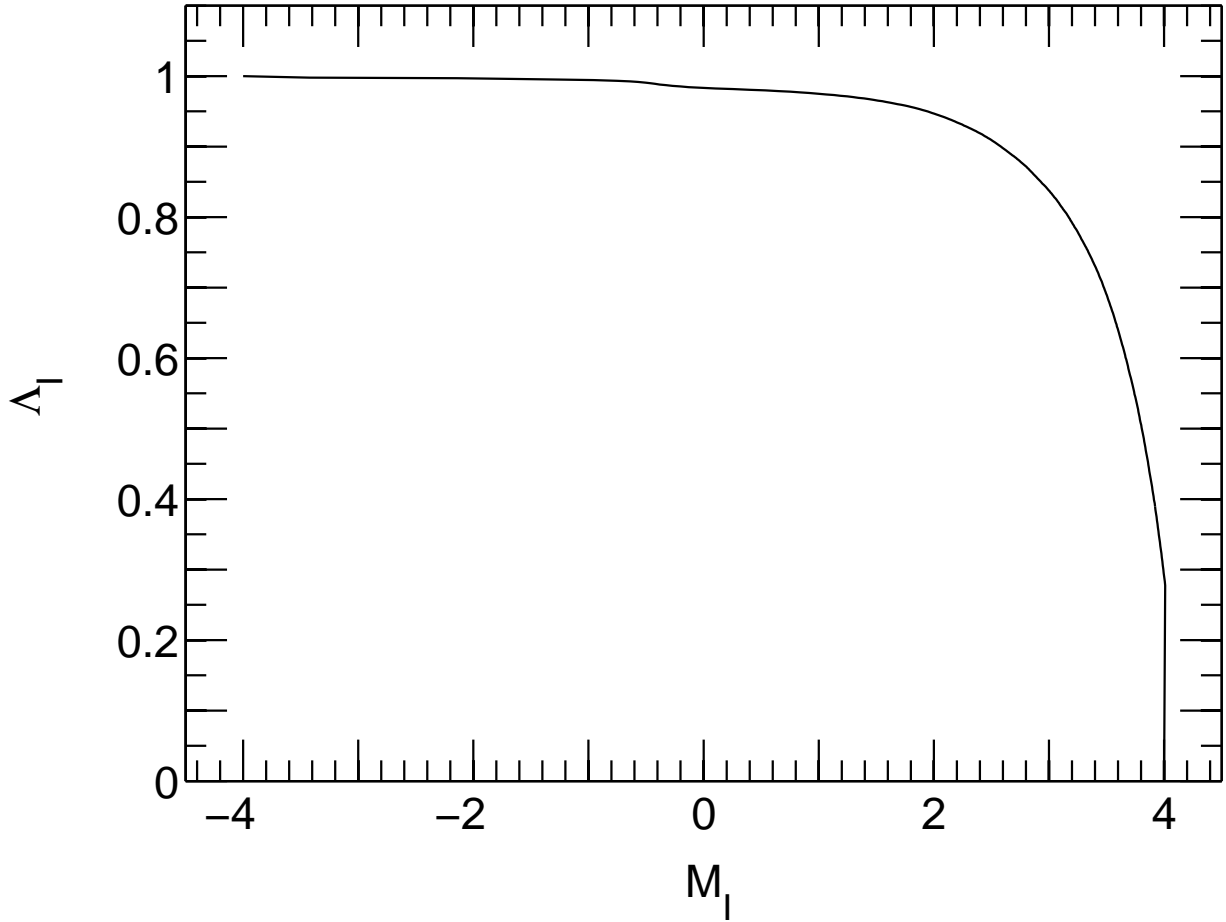}%
\figcaption[]{Function used to simulate incompleteness in mCMD and sCMD. The fraction of remaining stars ($\Lambda_I$) is given as a function of $I$. Values 1 and 0 mean that all or none stars remain, respectively.
\label{f10_comiacpop}}
\end{figure}

\begin{figure}[ht!]
\plotone{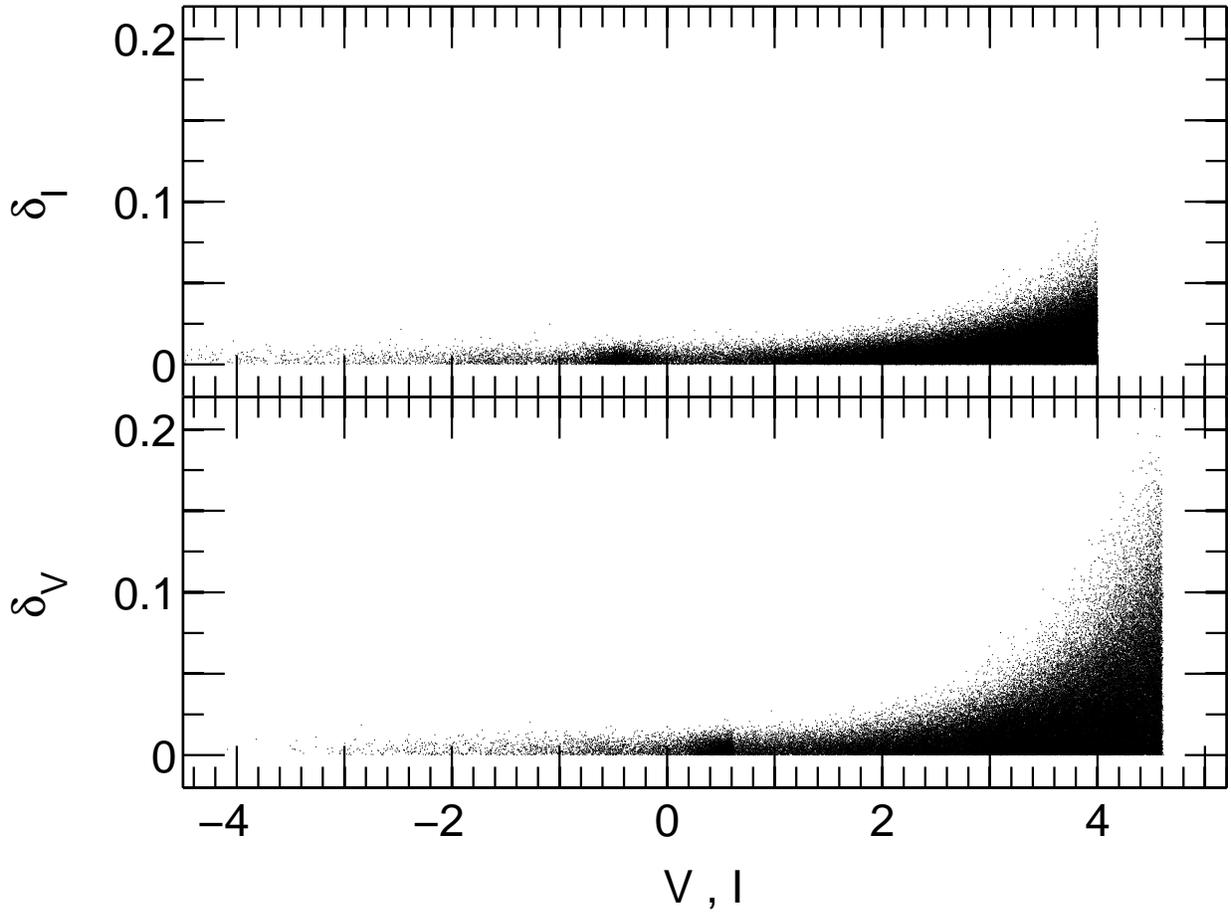}%
\figcaption[]{Shifts applied to the magnitudes in the mCMD and sCMD to simulate data dispersion.
\label{f11_gaussiacpop}}
\end{figure}

\begin{figure}[ht!]
\plotone{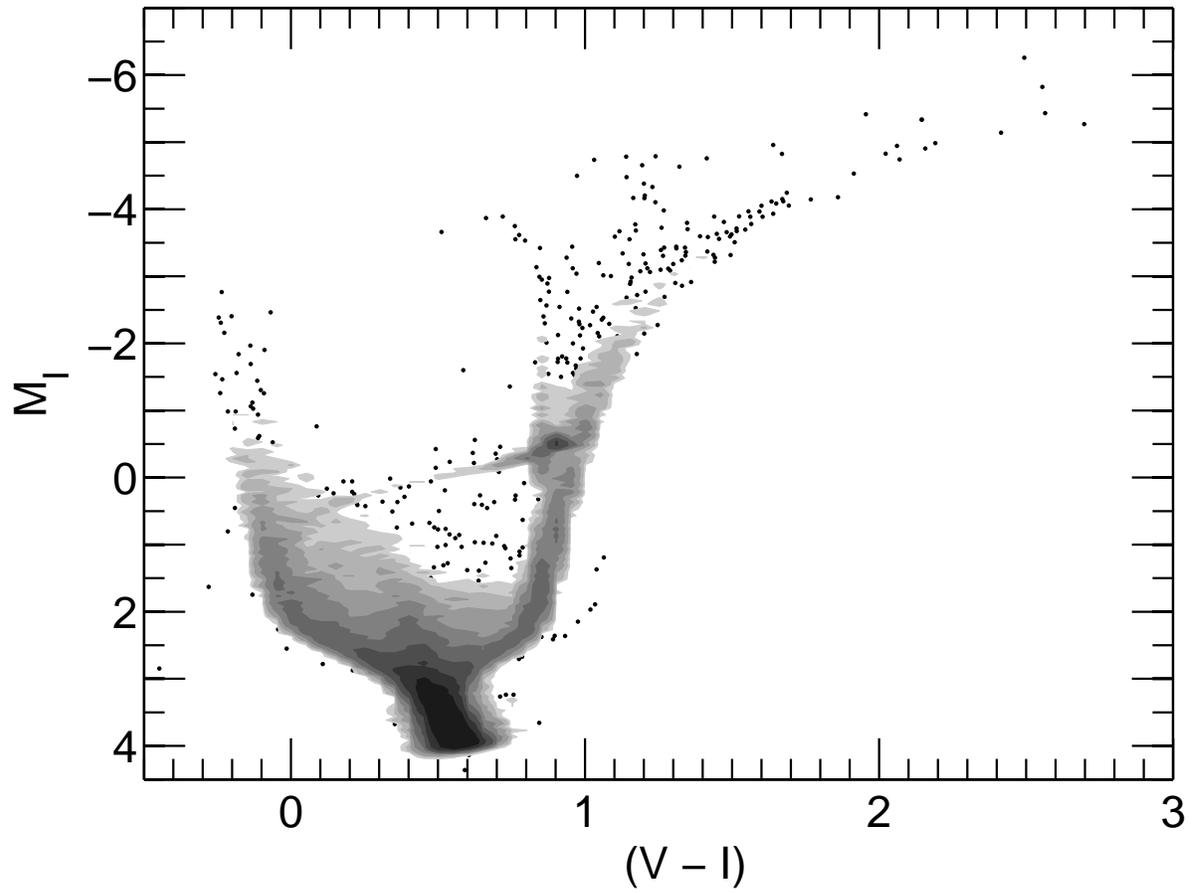}%
\figcaption[]{The CMD of the mock population after simulating observational errors in it. Caption is as for Fig. \ref{f02_dcm_noerr}.
\label{f12_dcm_err}}
\end{figure}

The best way to test how all these effects influence the SFH solution is working with synthetic CMDs, in which they have been simulated. For our particular purpose we have introduced random rejection and Gaussian dispersion of points, both depending on magnitude, in the mCMD and the sCMD. This approach is enough for this test although a more realistic simulation of observational effects should be used in real cases (see Hidalgo et al. 2009). For rejection we have used the completeness curve obtained for the CMD of the Phoenix dwarf galaxy (Hidalgo et al. 2009) and displayed in Fig. \ref{f10_comiacpop}. This curve shows the probability that a star of magnitude $I$ is conserved in mCMD or sCMD. A shift in $\delta V$ and $\delta I$ magnitudes is then applied to conserved stars. $\delta V$ and $\delta I$ are decided stochastically according to Gaussian distributions of $\sigma_V$ and $\sigma_I$ which are functions of $V$ and $I$ respectively and have been estimated using the CMD of the Phoenix dwarf galaxy (Hidalgo et al. 2009). Fig. \ref{f11_gaussiacpop} shows the $\delta V$ and $\delta I$ actually used. Figure \ref{f12_dcm_err} shows mCMD after simulation of the observational effects.

\placefigure{f10_comiacpop}
\placefigure{f11_gaussiacpop}
\placefigure{f12_dcm_err}

\begin{figure}[ht!]
\plotone{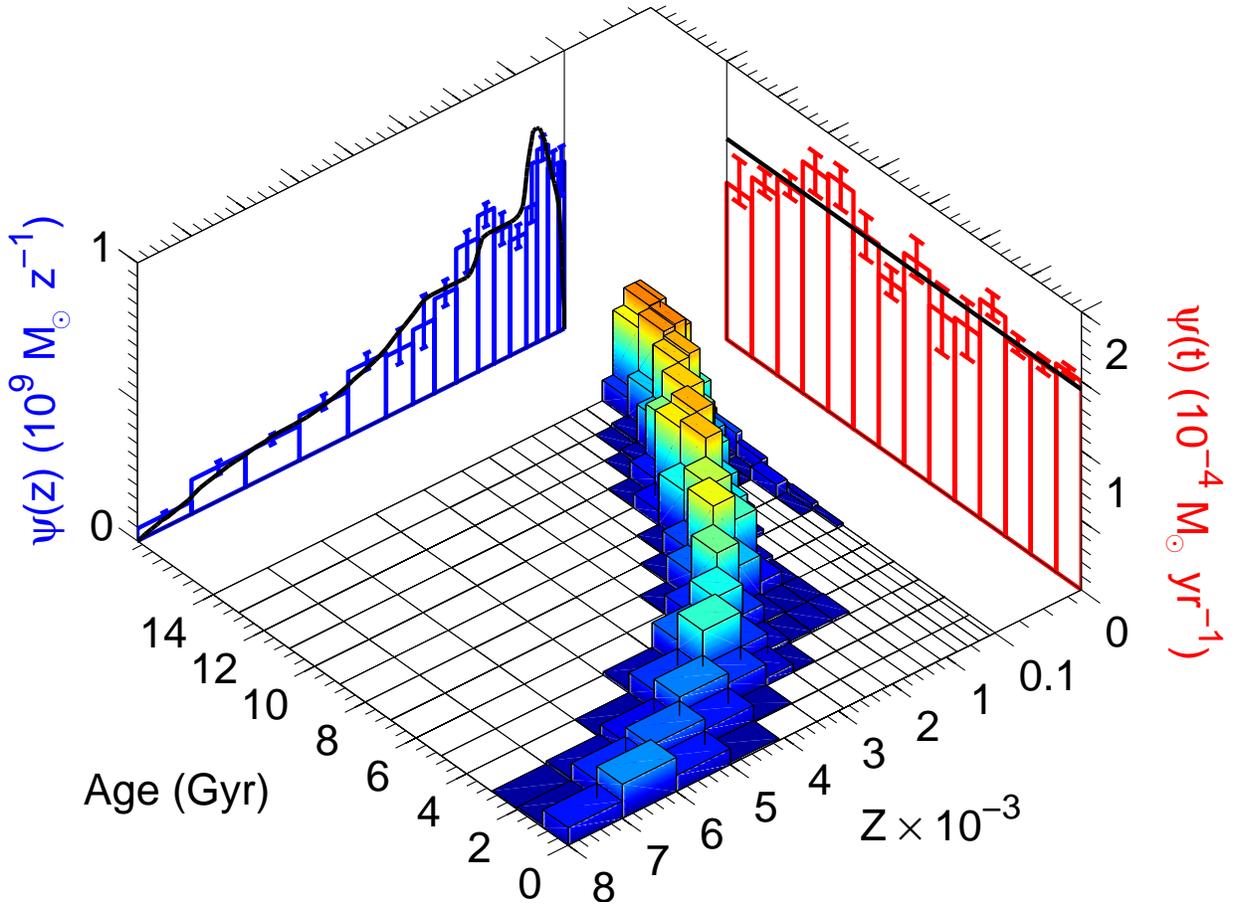}%
\figcaption[]{Solution of $\psi(t,z)$ obtained for the mock population with observational effects simulated. It is the average of 120 single solutions found using the {\it several solutions} procedure described in text. The input mono-dimensional $\psi(t)$ and $\psi(z)$ are shown by solid black lines. Error bars have been obtained applying the {\it several solutions} procedure. As a consequence of the observational effects, solutions appear noisier than for the observational effect free case shown in Fig. \ref{f05_nocrow}. 
\label{f13_crow}}
\end{figure}

\begin{figure}[ht!]
\plotone{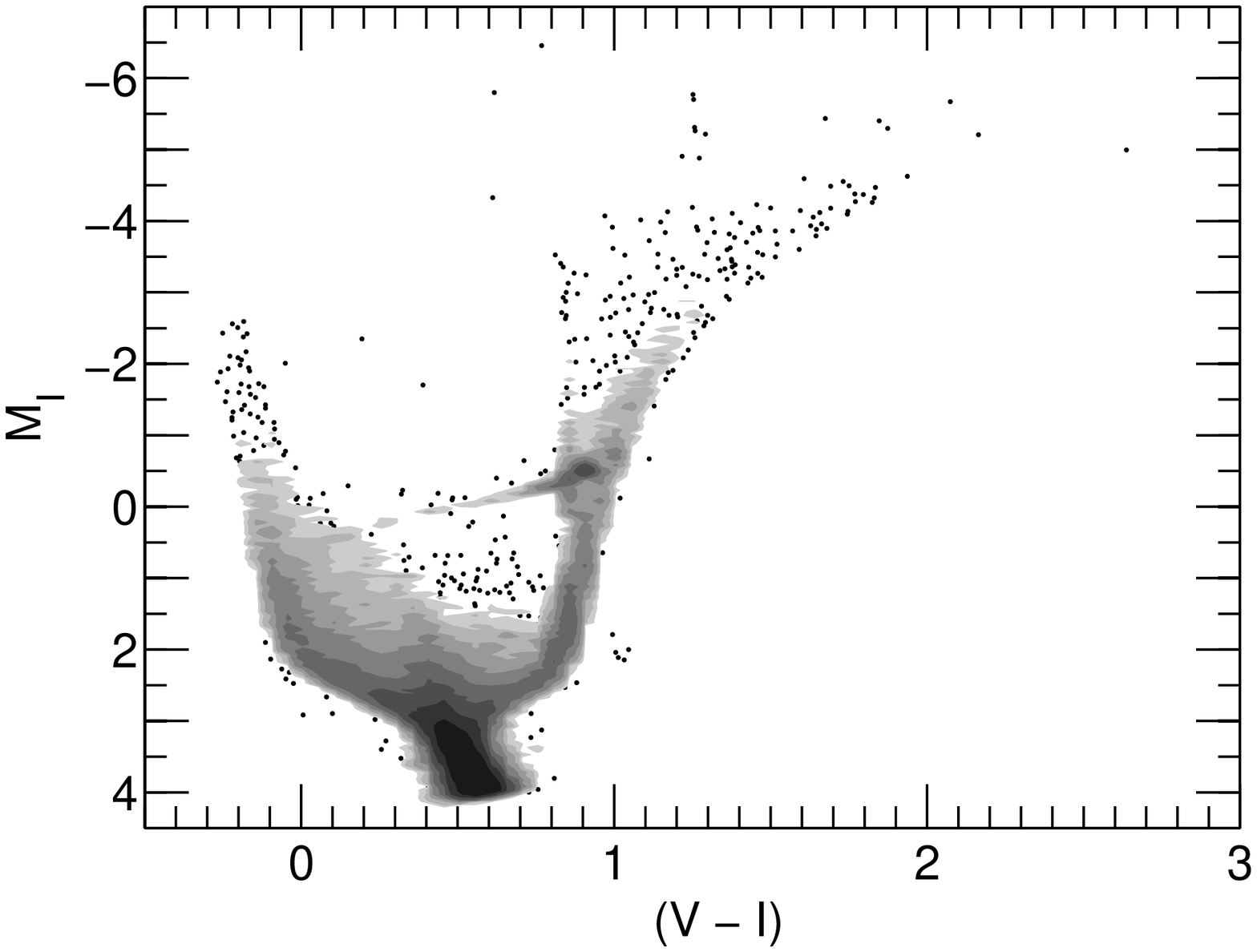}%
\figcaption[]{The CMD corresponding to the solution SFH shown in Fig.\ref{f13_crow}. To be compared with the mCMD shown in Fig. \ref{f12_dcm_err}.
\label{f14_sol_crow}}
\end{figure}

Figure \ref{f13_crow} shows the solution after simulating observational effects in mCMD and sCMD. Figure \ref{f14_sol_crow} shows the CMD corresponding to the SFH given in Fig. \ref{f13_crow}. As a result of the effects introduced in the CMD, $\psi(t,z)$ and also the projected $\psi(t)$ and $\psi(z)$ appear noisier than for the observational effect free case. However, results are similar to those of the observational effects free case, showing IAC-pop robustness against observational effects.

\placefigure{f13_crow}
\placefigure{f14_sol_crow}

\subsection{The effects of different stellar evolution libraries predictions}\label{librerias}

\begin{figure}[ht!]
\plotone{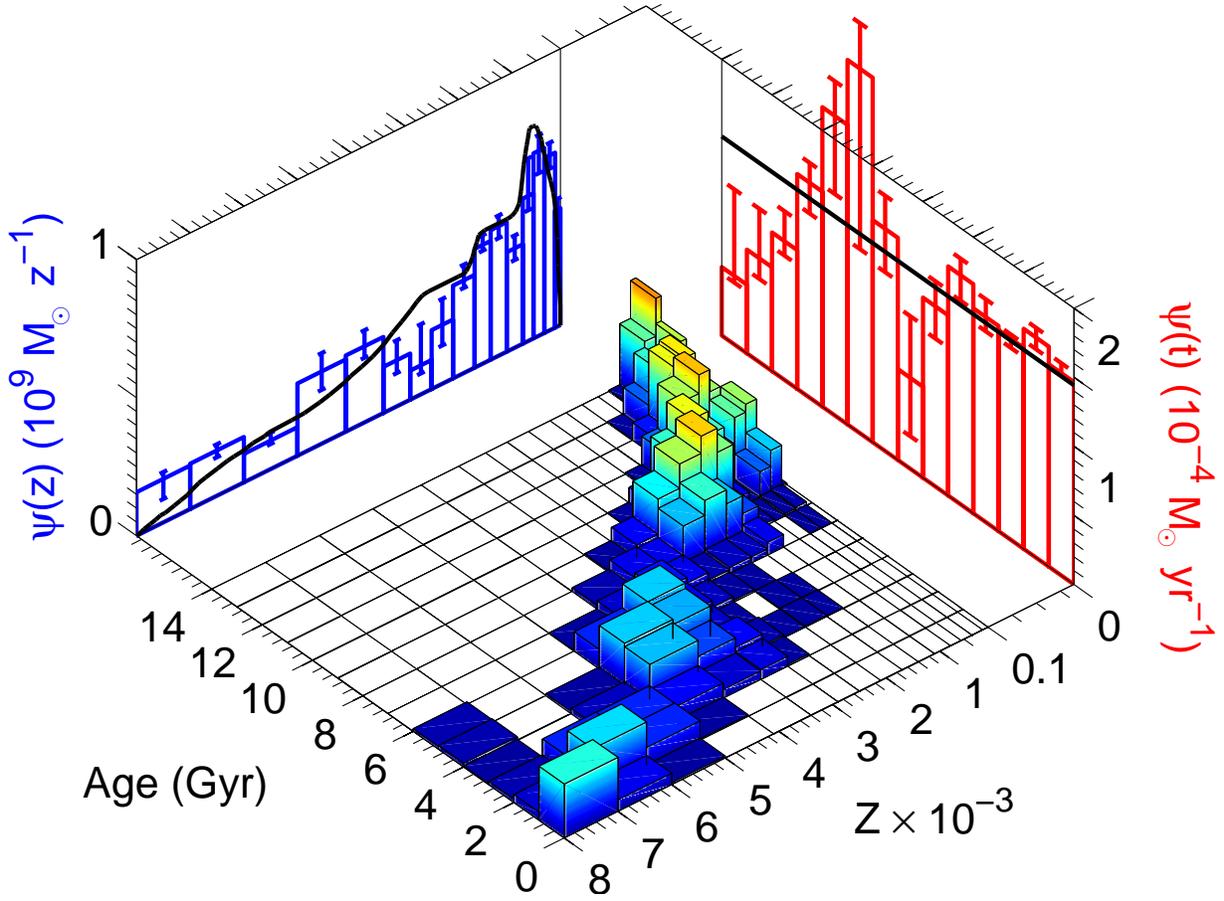}%
\figcaption[]{Solution using different stellar evolution libraries to compute sCMD and mCMD. It is the average of 120 single solutions found using the {\it several solutions} procedure described in text. The Teramo-BaSTI (Pietrinferni et al 2004) library is used to compute the mock population (mCMD) and the Padua one (Bertelli et al. 1994) for the global synthetic one (sCMD). Error bars have been obtained applying the {\it several solutions} procedure. 
\label{f15_nocrow_modelo2}}
\end{figure}

Working on CMDs in which observational effects have been simulated allows testing the capability of IAC-pop to handle real observations. However, sCMD and mCMD have been built with the same stellar evolution library, which removes the uncertainty introduced by our limited knowledge of stellar evolution. An idea of how this modeling affects results can be obtained by using different libraries to generate mCMD and sCMD. To this purpose, IAC-star has been used to compute an sCMD using the Padua stellar evolution library (Bertelli et al. 1994). The input parameters have been the same as before. Figure \ref{f15_nocrow_modelo2} shows the corresponding solution. Discrepancies between input and solution show now significant systematic effects. Differences are attributed to the stellar evolution model differences and therefore, the use of more than one stellar evolution library is recommended when analyzing a real population to test these effects.

\noindent Table \ref{consistable} summarizes the tests carried out in the present section. Column 1 identifies the test. Column 2 gives the $\chi^2_\nu$ value of the solution obtained for each test. Columns 3, 4 and 5 give, respectively, the total mass ($M_T$), mean age ($<age>$), and mean metallicity ($<z>$) of the stars in the mock population (first row) and the solutions. Agreement on integral and average values between mock population and solutions is good in all the cases, but $\chi^2_\nu$ is large for the case in which different stellar evolution libraries are used to compute the mock and the global synthetic populations.

\placefigure{f15_nocrow_modelo2}

\placetable{consistable}

\section{A further test: sharp bursts and time resolution}\label{bursts}

Time resolution is an important issue in the solution of the SFH. In general, it worsens for older ages. Moreover, it is ultimately limited by the quality of the data --which depends on the signal to noise with which each feature in the CMD is observed, the spatial resolution of the data and the number of stars in the CMD-- but not by the choice of the temporal sampling, which can be arbitrarily small. Olsen (1999), in his Figure 10, shows how the solution found for a synthetic stellar population is closer to the input as time sampling intervals become larger. This indicates that solutions averaged over large time intervals may be quite accurate, but that short time sampling could result in spuriously fluctuating solutions as shown also in Aparicio et al. (1997) (see also Skillman \& Gallart 2002).

\begin{figure}[ht!]
\plotone{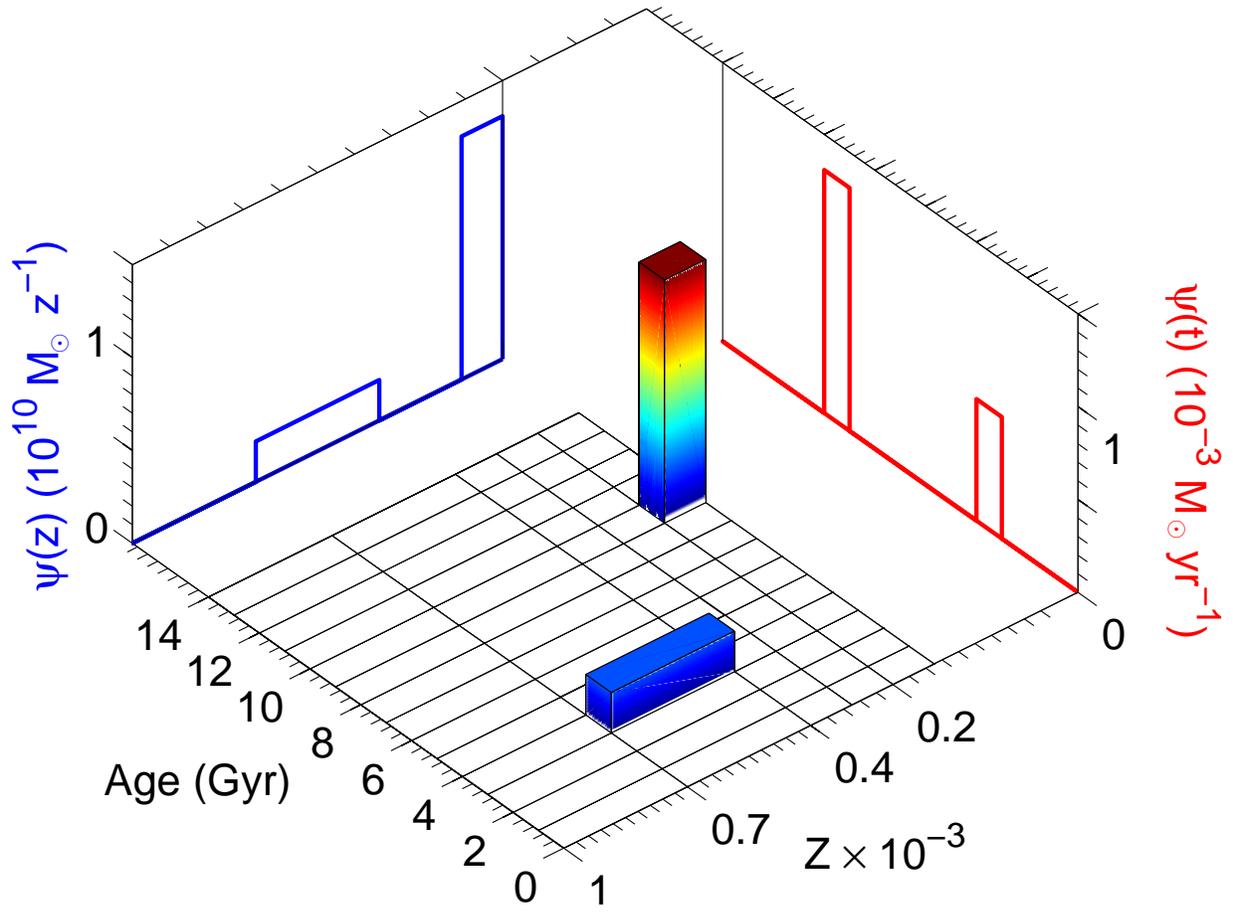}%
\figcaption[]{Mock stellar population composed by two narrow bursts used to test IAC-pop time resolution and precision. 
\label{f16_sfhmodelo_2burst}}
\end{figure}

\begin{figure}[ht!]
\plotone{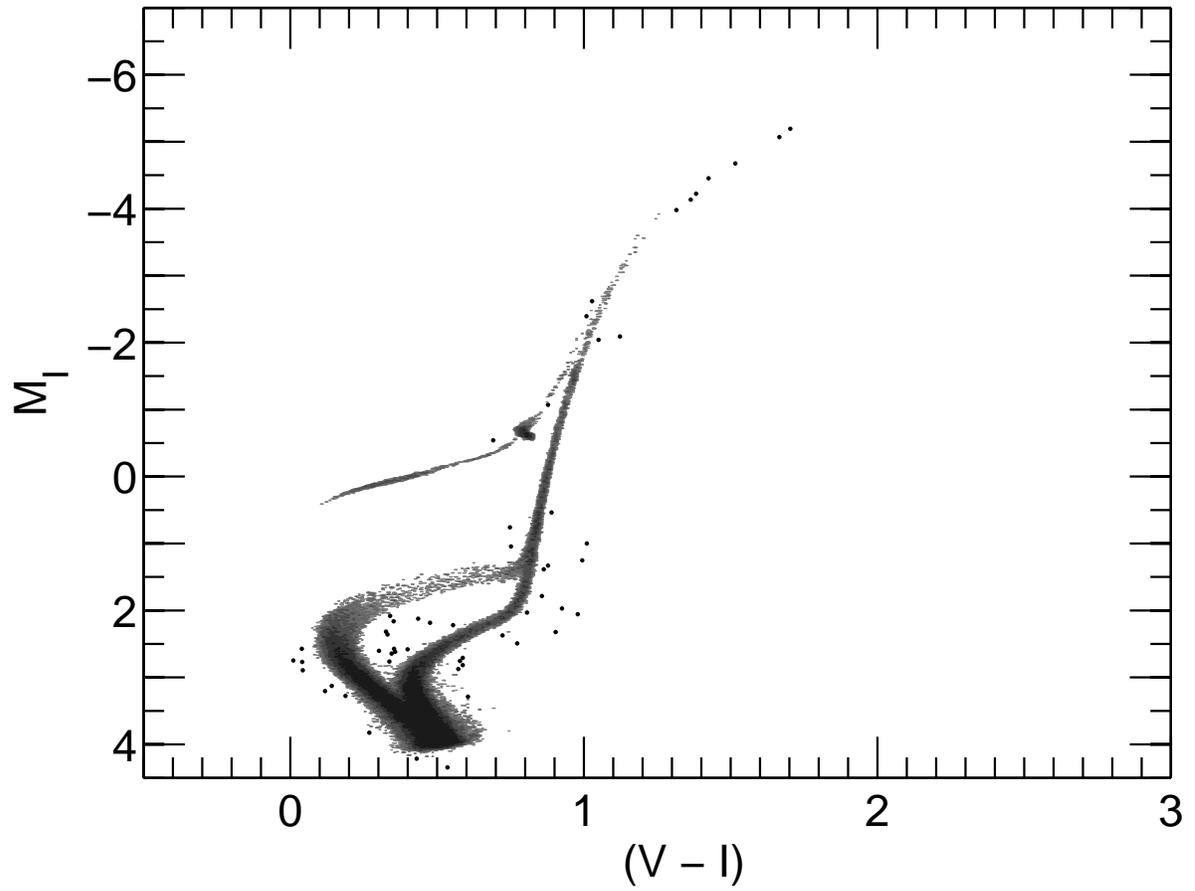}%
\figcaption[]{The mCMD corresponding to the SFH shown in Fig.\ref{f16_sfhmodelo_2burst}. Observational effects have been included.
\label{f17_dcm_err_2burst}}
\end{figure}

\begin{figure}[ht!]
\plotone{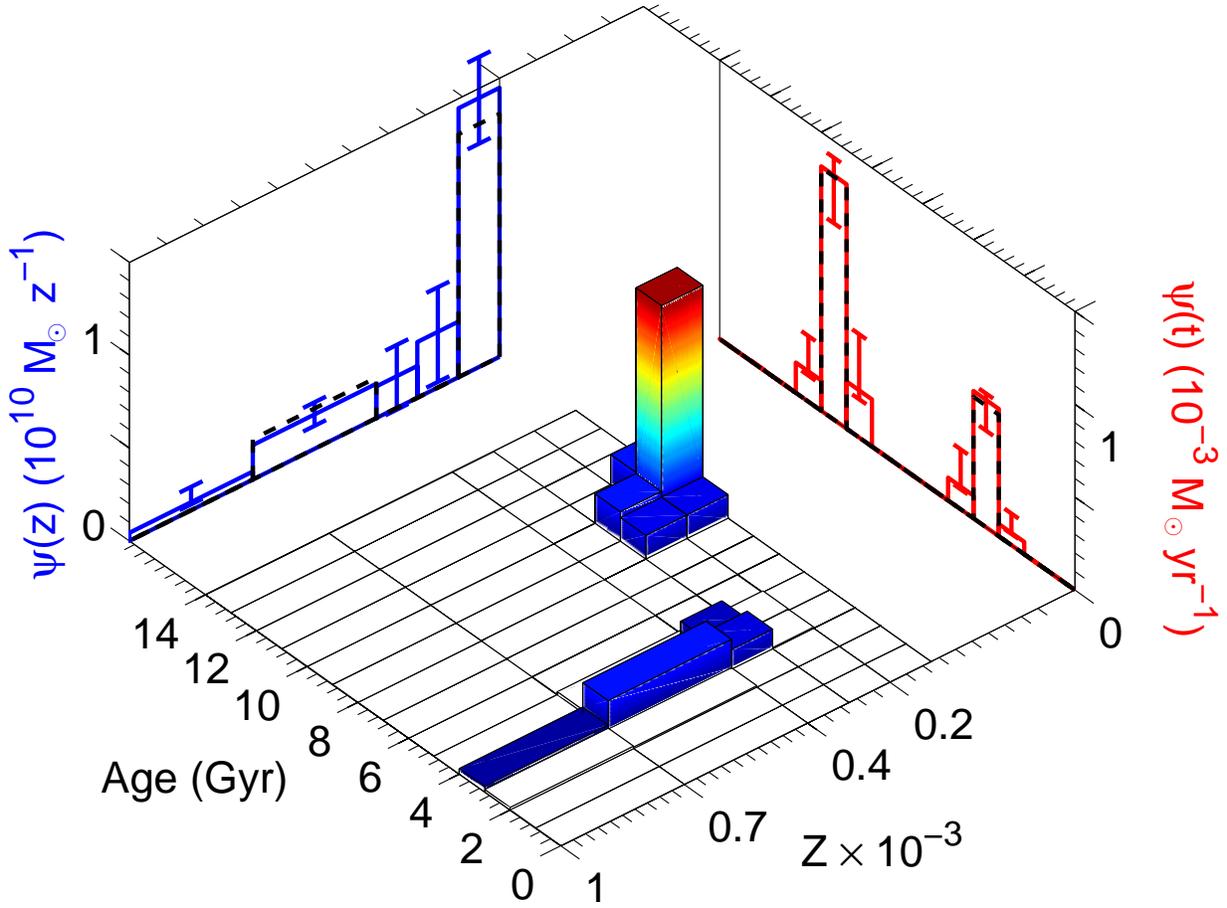}%
\figcaption[]{IAC-pop solution of $\psi(t,z)$ for the mock population shown in Fig. \ref{f16_sfhmodelo_2burst}. It is the average of 120 single solutions found using the {\it several solutions} procedure described in text. The Teramo-BaSTI (Pietrinferni et al 2004) library has been used to compute the mock and global synthetic populations. Dotted lines in the mono-dimensional representations $\psi(t)$ and $\psi(Z)$ show the input functions. Error bars have been obtained applying the {\it several solutions} procedure. 
\label{f18_sfh_gridcrow_2burst}}
\end{figure}

To test both time resolution and accuracy a mock population in which star formation has proceed in sharp bursts has been solved. To this purpose the mock population shown in Fig. \ref{f16_sfhmodelo_2burst} has been used together with the Teramo-BaSTI (Pietrinferni et al. 2004) library to compute the mCMD of Fig. \ref {f17_dcm_err_2burst} in which observational effects have been simulated. A solution has been sought using a sCMD computed from the global synthetic population shown in Fig. \ref{f03_sfhsintetica} and the Teramo-BaSTI library also, following the same procedure explained above. The solution is shown in Fig. \ref{f18_sfh_gridcrow_2burst}. The solution accuracy and precision are good, which provides a further test of internal consistency of IAC-pop, now including its capability to recover age and burst sharpness, at least for the time resolution and observational effects used in our example.

\placefigure{f16_sfhmodelo_2burst}
\placefigure{f17_dcm_err_2burst}
\placefigure{f18_sfh_gridcrow_2burst}

\section{Generalizing IAC-pop}\label{general}

IAC-pop has been designed with the idea in mind of solving for
$\psi(t,z)$. However, the code as it is, is blind to the nature of the
parameters behind the models used to generate the input information provided
to it. Rigorously, the code searches for the linear combination (with positive or null coefficients) of a set of $l$
reference vectors best reproducing a set of $k$ properties of a template
vector of dimension $k$. If run for solving the SFH, the latter is associated
to the stellar population, $k$ is the number of bins in which the CMD is
divided and $l=n\times m$ is the number of simple populations into which the
global synthetic sCMD has been divided. We have discussed the SFH case in detail, but it should be kept in mind that the code is of more
general application and that, if the $l$ reference
vectors are defined otherwise, the found solution will include other
properties of the galaxy or, in general, of the problem being analyzed.

\section{IAC-pop cook-book}\label{run}

IAC-pop is made available for free use. It can be downloaded from the
internet site {\tt http://iac-star.iac.es/iac-pop}. Together to IAC-pop, a
number of interactive software facilities will be also made available from this
site or from sites accessible from it, including the synthetic stellar populations and CMD generator
IAC-star. IAC-pop is currently offered compiled for several computer operating systems, like Linux, MacOS or Windows. In
the following we will describe the input parameters and summarize the content
of the output file.

\subsection{Input data}

The input is provided in two files. The first one, which we call the input parameter file, provides several choices
about the computation procedure, as follows:

\begin{itemize}

\item The input and output data files (see below).

\item The number of age and metallicity bins, $n$ and $m$. 

\item The number of boxes, $k$, defined in the observational and synthetic
  CMDs. 

\item A seed for the random number generator used by the genetic
  algorithm. 

\item Number of computed solutions, i.e. the number of times that
  the genetic algorithm is run. Each run starts with a different random number generator seed and provides an independent
  solution. In this way the possibility that the genetic code is trapped in a 
  secondary minimum is minimized (see Charbonneau 1995).

\item Number of generations computed for each single genetic solution.

\item Number of solutions changing input star count values according to a Poisson statistics. This provides a way of formal error estimation
  based upon the internal accuracy of the input observational data. This step is quite time consuming and should be set to 0 when not necessary. 

\item A good enough $\chi^2_{\nu}$ value. The program will accept as good enough a solution with a $\chi^2_{\nu}$ less or equal to this even if it has not reached the number of generations provided above. To switch-off this condition, set this value to 0.
 
\end{itemize}

The second file, which we call the input data file, is a list of $(l+1)\times k$ numbers. For more clarity we can assume them organized as $l+1$ rows
each one containing $k$ numbers, but it must be noted that separation into
different rows is not necessary. The content of the
rows is as follows:

\begin{itemize}

\item First row: observational data. The $k$ values provide the number of stars in each of the $k$ boxes defined in the CMD for the observational data. For a general problem, they provide the values of the evaluated property.

\item Rows from second to last (i.e., $(l+1)$): data for the simple synthetic populations. Explicitly, the $i$-th row provides the number of stars in each of the $k$ boxes defined in the CMD of the $(i-1)$-th simple population extracted from the global synthetic population. For a general problem, they provide values corresponding to the $(i-1)$-th simple model.

\end{itemize}

In other words, each row contains the star counts in the boxes defined in the
observational (first row) and the simple synthetic populations (second to last
rows) CMDs. Note that the $k$ boxes are the same for all the involved CMDs, observational and synthetic, and that the latter should contain a simulation of observational effects. The boxes set definition and the time and metallicity resolution are to be decided by the user. For
simplicity, in the following we will refer only to the case of interest in
this paper of solving a SFH using CMDs, but it should be kept in mind that
any other, similarly parametrized problem could be faced.

\subsection{Output file}

Upon completion, IAC-pop produces an output file containing the
solutions and the formal errors. The
content of the output file is structured in several labeled sets with the following content:

\begin{itemize}

\item The solutions obtained in the several independent code runs. Each row corresponds to one solution and contains $l+1$ data: the $l$ $\alpha_i$ parameters of the solutions plus its $\chi^2_\nu$ value.

\item The best of the former solutions, i.e., the solution having the $\chi^2_{\nu,min}$ value.

\item The average of the former solutions and their dispersions. This average could in some cases be preferable to the solution having the $\chi^2_{\nu,min}$ value as it smooth out possible no realistic fluctuations. The dispersions are indicative of the solution stability, but not of its actual error.

\item Dispersions of the computed solutions around the best one and the number of solutions found above and below the best one. This information is complementary to evaluate solution stability and fluctuations. 

\item Formal error estimate. Solutions found for several observational input data sets in which the star counts corresponding to each of the $k$ boxes defined in the CMD are randomly changed according to a Poisson statistics within $\sqrt n_j$, where $n_j$ is the actual number of stars in box $j$. It must be noted that these errors are underestimates of the total internal errors.

\item Standard deviations of the former solutions. Upward and downward error bars and number of solutions above and below the best one are obtained in the same way as above.

\end {itemize}

\section{Final remarks and conclusions}\label{conclusions}

Summarizing, IAC-pop is a program designed to solve the SFH of a
complex stellar population system, like a galaxy, from the analysis of the
CMD. To this purpose, IAC-pop uses a genetic algorithm (Charbonneau 1995)
to minimize a reduced Mighell's, $\chi_\nu^2$, merit function (Mighell 1999) obtained from comparison of the parametrization of an observed and a synthetic CMD. The code main characteristics can be sketched as follows:

\begin{itemize}

\item The code needs the computation of only a single global synthetic
  CMD. As many simple population model CMD as necessary are later extracted from  it. We call sCMD this initial global synthetic CMD.

\item It is designed to solve simultaneously for age and metallicity
  distributions; i.e. for the star formation rate as a function of time and metallicity, which provides also the chemical enrichment 
  law.

\item The parametrization of observed and synthetic CMDs
  is done by dividing the CMDs in several boxes and counting out
  the stars in each one. This is the information provided to the code. 

\item The former implies that the code application is not restricted to solve 
  the problem of the SFH, but it is of general application to problems in
  which a similar parametrization can be done. 
  
\item It is important to note that observational effects (crowding, blending,
  completeness, etc) must be simulated in the synthetic global data or in the input parametrization prior to run IAC-pop. 

\item A genetic algorithm is used to
  minimize the aforementioned merit function $\chi^2_\nu$.

\item The final solution is provided as a linear combination of positive or null coefficients of the input
  simple population models. 

\item In its current version, IAC-pop provides the best solution for one or several runs. It also provides a formal error estimate based on Poissonian random fluctuations of the input. It should be noted that these formal errors seem to be underestimates of the total internal errors of the problem. More realistic error estimates are obtained as the dispersion of several solutions obtained with several input parametrization to IAC-pop (which we have named {\it several solutions} procedure). The IAC-pop user is strongly encouraged to implement it. The routine MinnIAC will be provided under request to help in this job. 

\end{itemize}

IAC-pop has been run through several consistency tests. To this purpose a mock stellar population has been computed using IAC-star and analyzed with IAC-pop to obtain its SFH as if it were a real one. Results have been compared with the input used to compute the mock population under different assumptions. 

\begin{itemize}

\item For the first test, a SFH continuously varying as a function of time and metallicity was used for the mock population. The test simply consisted in deriving the SFH in an observational error free scenario and using a global sCMD computed with the same stellar evolution library as for the mock population. Results were in quite good agreement with input, proving the IAC-pop code internal consistency. 

\item The second test was a repetition of the former but after simulating observational effects both in the mock population mCMD and the global sCMD. Although noisier, the resulting SFH was also in good agreement with the input SFH, showing the robustness of the method against realistic observational effects. 

\item The third test was a repetition of the former, including observational effects, but using different stellar evolution libraries to compute the mock population mCMD and the global sCMD. This is expected to reproduce the effects introduced in the solution by the inaccurate knowledge of the stellar evolution physics. Systematic trends show up due to the differences in the stellar evolution models and therefore, the use of more than one stellar evolution library is recommended when analyzing a real population to control these effects.

\item Finally, the fourth test is done on a mock population made of two sharp bursts at young and intermediate to old ages. The bursts are well reproduced, even if observational effects are simulated, if the same stellar evolution library is used to compute the mock population mCMD and the global sCMD. 

\end{itemize}

In summary, IAC-pop has been shown to be an useful tool to obtain the SFH from the CMD of resolved stellar systems. The program can be downloaded from the site
{\tt http://iac-star.iac.es/iac-pop}, with the only requirement of referencing this paper and acknowledging the IAC in any derived publication. The routine MinnIAC is also offered under request and at user's risk. Its use requires also referencing this paper. It is intended to
produce further improved versions of the program after feed-back by the user
community.

\acknowledgments

Developing IAC-pop and MinnIAC has been greatly benefited from long discussions maintained with and many tests performed by S. Cassisi, C. Gallart, M. Monnelli and E. Skillman and by the very interesting comments of the anonnymous referee. 
The authors are funded by the IAC
(grant 310394) and by the Science and Technology Ministry of the Kingdom
of Spain (grants AYA2004-3E4104 and AYA2007-3E3507).

\clearpage

\clearpage

\clearpage

\begin{table}[h!]
\begin{center}
\caption{\label{consistable} Results for the self-consistency test}
\scriptsize{
\begin{tabular}{l c c c c c}
\noalign{\smallskip}
\hline\hline
\noalign{\smallskip}
CMD                 &$\chi^2_\nu$ &$M_T$ ($10^6 M_\odot$) &$<age>$ (Gyr) &$<z>$\\
\noalign{\smallskip}
\hline
\noalign{\smallskip}
Mock CMD             &                &2.02                   &7.00           &0.0026\\
No observ. effects             &1.1             &$2.00\pm 0.02$         &$6.9\pm 1.1$ &$0.0026\pm 0.0005$\\
With observ. effects                &0.9             &$2.00\pm 0.02$         &$6.9\pm 1.1$ &$0.0026\pm 0.0005$\\
BaSTI-Padua          &4.6             &$2.03\pm 0.03$         &$6.7\pm 1.0$ &$0.0030\pm 0.0006$\\
\noalign{\smallskip}
\hline
\hline
\end{tabular}
}
\normalsize
\rm
\end{center}
\end{table}


\end{document}